\documentclass[reprint,superscriptaddress,amsmath,amssymb,aps,prl,longbibliography]{revtex4-2}
\usepackage[utf8]{inputenc}
\usepackage{dcolumn}
\usepackage{bm}
\usepackage{graphicx}
\usepackage{amsmath}
\usepackage{amssymb}
\usepackage{placeins}
\usepackage{float}
\usepackage{color, xcolor}
\usepackage[normalem]{ulem}
\usepackage{newunicodechar}
\newunicodechar{）}{)}
\usepackage{soul}
\soulregister\cite7
\soulregister\ref7

\begin{document} 

\title{Evolution of magnetic correlation in doped Hubbard model with altermagnetic spin splitting}

\author{Yinlong Li}
\thanks{These authors contributed equally.}
\affiliation{School of Science, Harbin Institute of Technology, Shenzhen, 518055, China}
\affiliation{Shenzhen Key Laboratory of Advanced Functional Carbon Materials Research and Comprehensive Application, Shenzhen 518055, China.}

\author{Rana Imran Mushtaq}
\thanks{These authors contributed equally.}
\affiliation{School of Science, Harbin Institute of Technology, Shenzhen, 518055, China}
\affiliation{Shenzhen Key Laboratory of Advanced Functional Carbon Materials Research and Comprehensive Application, Shenzhen 518055, China.}

\author{Ji Liu}
\affiliation{School of Science, Harbin Institute of Technology, Shenzhen, 518055, China}
\affiliation{Shenzhen Key Laboratory of Advanced Functional Carbon Materials Research and Comprehensive Application, Shenzhen 518055, 
China.}

\author{Wing Chi Yu}
\email{wingcyu@cityu.edu.hk}
\affiliation{
Department of Physics, City University of Hong Kong, Kowloon, Hong Kong
}

\author{Xiaosen Yang}
\email{yangxs@ujs.edu.cn}
\affiliation{Department of Physics, Jiangsu University, Zhenjiang, 212013, China.}

\author{Cho-Tung Yip}
\affiliation{School of Science, Harbin Institute of Technology, Shenzhen, 518055, China}

\author{Ho-Kin Tang}
\email{denghaojian@hit.edu.cn}
\affiliation{School of Science, Harbin Institute of Technology, Shenzhen, 518055, China}
\affiliation{Shenzhen Key Laboratory of Advanced Functional Carbon Materials Research and Comprehensive Application, Shenzhen 518055, China.}


\begin{abstract}

The evolution of magnetic correlation in strongly correlated electron systems with altermagentic spin splitting remains largely unexplored. Here we investigate how spin splitting generated by spin-dependent next-nearest-neighbor hopping $t'$ reshapes the Fermi surface nesting and van Hove singularities in the two-dimensional square-lattice Hubbard model, leading evolution of magnetic instabilities. Using the constrained-path quantum Monte Carlo method, we find the dominant magnetic correlation as functions of the filling and $t'$ by computing the momentum-resolved spin structure factor. The analysis reveals a transition from antiferromagnetic $(\pi,\pi)$ order in the isotropic, half-filled system to non-collinear spiral  $(\pi,q)$ order upon increasing the spin-dependent anisotropy or doping away from half-filling, ultimately entering a short-range correlation regime where stripe and spiral correlation coexist. These findings highlight a possible route to realizing spiral correlation in altermagnetic systems, potentially providing a platform for spintronic devices that exploit non-collinear spin textures.
\end{abstract}

\maketitle

\section{\uppercase\expandafter{\romannumeral 1}. INTRODUCTION}
\label{Introduction}

In strongly correlated electron systems, the interplay of doping, lattice geometry, and electronic correlation engenders a rich landscape of competing magnetic phases, exemplified by the evolution from antiferromagnetism to complex spin textures in materials such as cuprates~\cite{bednorz1986possible, goremykin2024antiferromagnetic, wagman2013two, Kastner1998magnetic}. Theoretically, this magnetic reconstruction is intrinsically governed by the structure of the Fermi surface and the associated particle–hole nesting instability~\cite{he2019fermi, khokhlov2020dynamical}. Specifically, when extensive regions of the Fermi surface are connected by a magnetic wave vector, the spin structure factor exhibits a pronounced peak, promoting the formation of magnetic correlation \cite{igoshev2010incommensurate}. As carrier concentration varies, the consequent deformation of the Fermi surface shifts the optimal nesting vector from the $(\pi, \pi)$ point to spiral values, thereby driving the transition between distinct magnetic ground states, such as spiral order~\cite{fujita2002static, yamada1998doping, markiewicz2024theory, kang2019evolution}.

\begin{figure}[b!]
    \centering
    \includegraphics[width=0.5\textwidth]{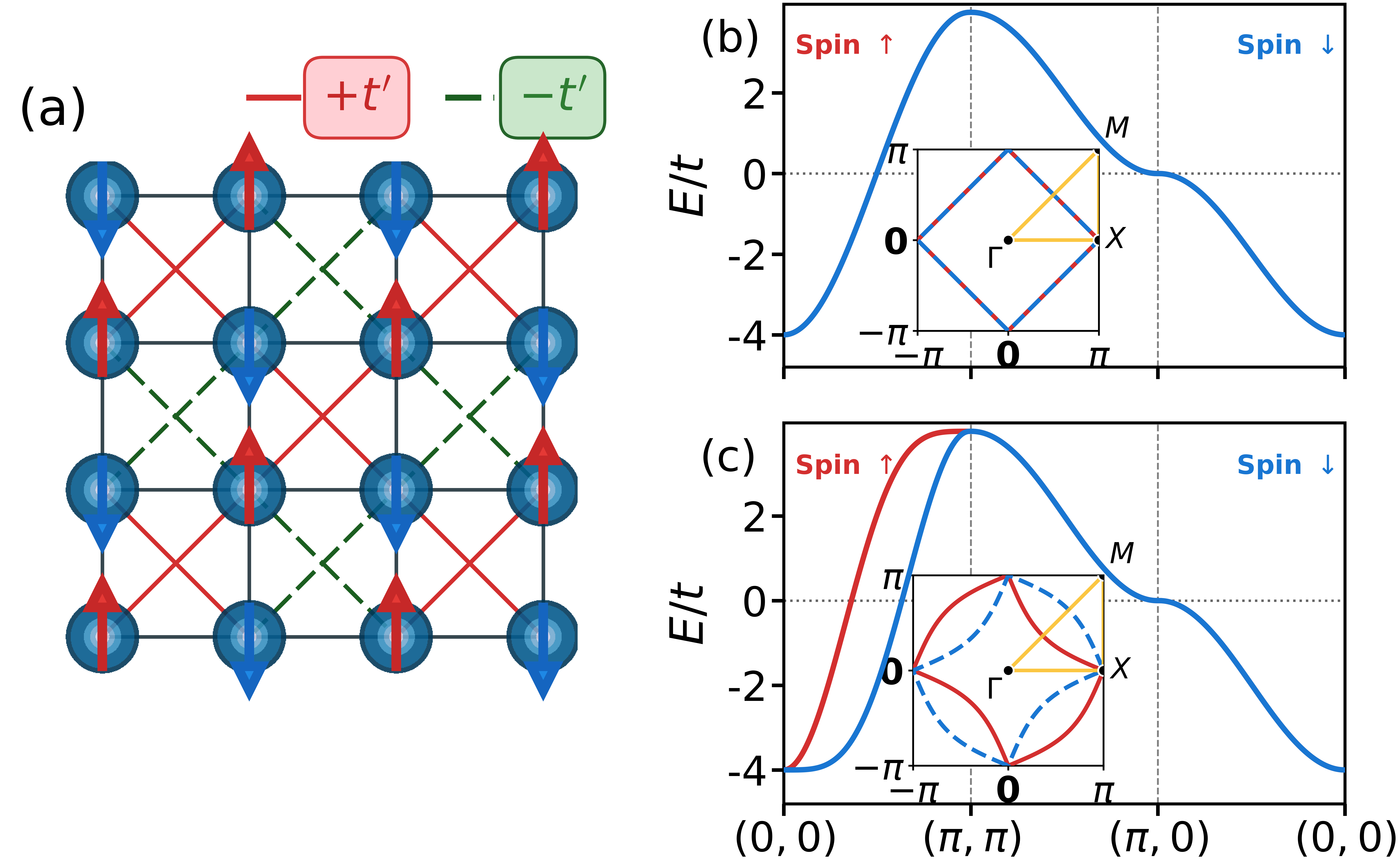}
    \caption{
Schematic representation of the spin-anisotropic $t'$ Hubbard model and its resulting band splitting. 
(a) Square lattice with on-site interaction $U$ and sign-alternating next-nearest-neighbor (NNN) hopping $\pm t'$, giving opposite effective NNN hopping for the two spin species while maintaining zero net magnetization.
(b) Noninteracting band dispersion along the high-symmetry path of the Brillouin zone and the corresponding spin-resolved Fermi surface (inset) for $t' = 0$, where the spectrum is spin degenerate.
(c) Noninteracting band dispersion and spin-resolved Fermi surface for $t' = 0.5$, where the term $4t'\sin k_x \sin k_y$ induces momentum-dependent spin splitting and distinct Fermi surface contours for spin up (red) and spin down (blue).}
    \label{fig:my_label1}
\end{figure}

To explore these phenomena theoretically, the two-dimensional Hubbard model is widely recognised as the standard starting point for studying doped Mott insulators \cite{dong2022mechanism, scalapino2012common}. Despite its formal simplicity—comprising only kinetic terms describing electronic transitions and a local Coulomb repulsion term $U$—this model successfully captures the essence of strongly correlated physics: the competition between kinetic energy and strong electron interactions\cite{arovas2022hubbard, qin2022hubbard}. It is precisely this competition that gives rise to a variety of intertwined and competing quantum phases, including antiferromagnetism and charge ordering \cite{beenen1995superconductivity}. Consequently, the Hubbard model has become an indispensable theoretical tool for studying how strong correlation reshape competing particle–hole instabilities across doping and band-structure tuning \cite{dong2022mechanism}. However, the minimal Hubbard model containing only nearest-neighbor (NN) hopping ($t$) exhibits particle-hole symmetry at half-filling, which contradicts the experimentally observed significant asymmetry in the electron-doping versus hole-doping phase diagrams of copper oxides \cite{rashid2024effect}. 

\textcolor{black}{The introduction of the next-nearest-neighbor (NNN) hopping term $t'$ to construct the so-called $t-t'$ Hubbard model has important physical consequences \cite{rashid2024effect}. In the conventional $t-t'$ Hubbard model, the NNN hopping $t'$ is typically spin-independent. Consequently, in the absence of spontaneous magnetic ordering, the single-particle band structure remains spin-degenerate, with magnetism primarily determined by the interplay between interactions and the geometry of the Fermi surface. In contrast, the $t'$ term considered here adopts a spin-dependent and diagonally anisotropic form, inducing spin splitting in momentum space at the single-particle level. However, as the two spin splitting exhibit a compensatory relationship, the system retains magnetically compensated and has zero net magnetization. The non-zero $t'$ term explicitly breaks particle-hole symmetry, rendering the Fermi surface more realistic, while profoundly affecting the nesting condition \cite{cade2020strategies}. Consequently, $t'$ plays a decisive role in regulating the competition between different ground states—particularly in the contest between antiferromagnetic and spiral magnetic ordering \cite{beenen1995superconductivity}—and exerts a critical influence on the evolution and distribution of magnetic order discussed herein \cite{markiewicz2023high}.} 

Although the NNN hopping $t'$ is known to reshape the electronic structure by shifting van Hove singularitie (VHS) and redistributing low-energy intensity in momentum space~\cite{cade2020strategies, markiewicz2023high}, its precise role in driving the evolution of magnetic correlation remains unclear. This unresolved issue is central to understanding how competing magnetic instabilities emerge in the doped Hubbard model, where modest changes in the Fermi surface can qualitatively alter the dominant ordering tendencies.  Previous studies indicate that $t'$ can influence the magnetic order, resulting in the formation of stripe or non-conjugate magnetic correlation~\cite{markiewicz2024theory}. A sufficiently large  $t'$ has been argued to stabilize a single spiral magnetic correlation over a broad range of carrier concentrations~\cite{beenen1995superconductivity}, whereas small $t'$ or increased doping typically produces a more intricate structure of the bare susceptibility, leading to near-degeneracy among multiple competing ordering vector and rendering the phase competition highly sensitive to microscopic details.

\textcolor{black}{Recent work on altermagnetism shows that a material can be collinear and magnetically compensated while still exhibiting a clear momentum-dependent spin splitting due to symmetry~\cite{PhysRevX.12.031042}. Inspired by this idea, we use a minimal square-lattice Hubbard model where the NNN hopping $t'$ depends on spin and direction, which produces a $d$-wave-like spin splitting already in the noninteracting band structure while keeping zero net magnetization. With the constrained-path quantum Monte Carlo (CPQMC) at finite $U$, we then track how the leading magnetic correlations (N\'eel/stripe/spiral) and their coherence evolve across the $(n,t')$ plane, in contrast to the usual spin-independent $t$--$t'$ model.}

\textcolor{black}{Recent weak-coupling renormalization-group studies have investigated closely related altermagnetic-inspired square-lattice Hubbard models with small spin splitting. For example, Parthenios et al.~\cite{PartheniosPRB2025} and Rao et al.~\cite{RaoPRB2025} analyzed interaction-driven instabilities in such models using renormalization-group techniques. To complement these studies, we use constrained-path quantum Monte Carlo at finite $U$ to directly compute the spin structure factor and real-space spin correlations over a broad $(n,t')$ scan, allowing us to map how N\'eel, stripe, and spiral correlations compete and evolve beyond weak-coupling instability analysis.}

The remainder of this paper is organized as follows: Section 2 outlines the theoretical framework and computational methodology; Section 3 presents our numerical results and detailed analysis; and Section 4 provides a conclusion and discussion of our findings.


\section{\uppercase\expandafter{\romannumeral 2}. Model and Method }
\label{Model and Method}
\subsection{A. Next-nearest-neighbor Hubbard model}
\label{NNN model}
The Hamiltonian of the Hubbard model is
 
\begin{equation}
    \begin{aligned}
\hat{H}= &\Big [- t \sum_{\langle i j\rangle, \sigma}  \hat{c}_{i \sigma}^{\dagger} \hat{c}_{j \sigma}- \sum_{i,\sigma} \Big ( t^{\prime}_{\hat x +\hat y,\sigma}\hat c^\dagger_{i\sigma}\hat c_{i+\hat x+\hat y,\sigma}\\
 &\textcolor{black}{+} t^{\prime}_{\hat x -\hat y,\sigma} \hat c^\dagger_{i\sigma}\hat c_{i+\hat x-\hat y,\sigma}\Big) 
   +   h.c. \Big] 
         +U \sum_i \hat{n}_{i \uparrow} \hat{n}_{i \downarrow}-\mu \sum_{i \sigma} \hat{n}_{i \sigma},
    \end{aligned}
\label{eq:Hubbard_model}
\end{equation}
where $i$ and $j$ label lattice sites, $\hat{c}^{\dagger}_{i\sigma}$ ($\hat{c}_{i\sigma}$) creates (annihilates) an electron with spin $\sigma\in\{\uparrow,\downarrow\}$ at site $i$, and $\hat{n}_{i\sigma}=\hat{c}^{\dagger}_{i\sigma}\hat{c}_{i\sigma}$ is the local number operator. The first term describes NN hopping on the square lattice, where $\langle ij\rangle$ denotes NN bonds. We set the NN hopping $t$ as the energy unit, with $t=1$. For the NNN directions, for different spins, the $t^{\prime}$ term carries different signs in different directions, with values ranging from [0.0, 1.0] (intervals of 0.1), for up spin, $t^{\prime}_{x+y,\uparrow} = t^{\prime}$ and $t^{\prime}_{x-y,\uparrow} = - t^{\prime}$; for down spin, $t^{\prime}_{x+y,\downarrow} = - t^{\prime}$, $t^{\prime}_{x-y,\downarrow} = t^{\prime}$. The spin-dependent NNN hopping generates a momentum-dependent spin splitting already at the single-particle level, while maintaining a vanishing net magnetization in the absence of interactions, as shown in Fig.~\ref{fig:my_label1}. \textcolor{black}{We emphasize that Eq.\ref{eq:Hubbard_model} is introduced as a minimal effective lattice model for a compensated, momentum-dependent $d$-wave-like spin splitting on the square lattice. The specific sign structure of the spin-conserving NNN hopping is chosen as the minimum effective form that produces such spin splitting while preserving zero net magnetization and avoiding additional parameters. In this sense, $t'$ should be viewed as an effective symmetry-motivated band-splitting term inspired by altermagnetic phenomenology, rather than a material-specific Wannier Hamiltonian or conventional relativistic spin--orbit coupling. More elaborate extensions with additional hopping anisotropies are possible, but they are beyond the scope of the present minimal study.}

The third term is the on-site Hubbard interaction $U$, and $\mu$ is the chemical potential. We set the on-site interaction $U=4$, which places the system in the intermediate coupling regime, where the interplay between kinetic energy and electron-electron interactions is most pronounced. The interaction is sufficient to induce robust magnetic correlation and potential ordering instabilities without fully localizing the charge carriers, thereby allowing for a detailed examination of the competition between Fermi surface nesting effects and strong correlation physics.



We decompose the Hamiltonian into a kinetic (noninteracting) part, $H_K$, and an interaction part, $H_N$. The noninteracting term $H_K$ is diagonalized in momentum space, yielding the noninteracting band structure. Its explicit form is given in Eq.~\eqref{eq:energy function}.
\begin{equation}
\varepsilon (\mathbf{k}) = -2t\bigl (\cos k_x + \cos k_y\bigr) \pm 4 t' \sin k_x \sin k_y + \mu \, .
\label{eq:energy function}
\end{equation}
The altermagnetic term $t^{\prime}$ significantly changes the shape of Fermi surface. Fig.~\ref{fig:my_label1} schematically summarizes the Hubbard model considered in this work and the spin splitting induced by the NNN hopping $t'$. 

To further investigate the spin splitting phenomenon in magnetism, we introduce the total momentum domain spin polarization $\Delta _{tot}$:

\begin{equation}
\Delta _{tot}= \sum_{k}|\Delta n (k)| = \sum_{k}|n_{\uparrow} (k) - n_{\downarrow} (k)|.
\end{equation}
Here, $\Delta n (k)$ represents the spin polarization in momentum space. A nonzero value of this quantity reveals the presence of momentum-dependent spin splitting, which breaks the local spin degeneracy in reciprocal space despite the system exhibiting zero net magnetization globally.
Through a comprehensive scan of the $(t^{\prime}/t, n)$ parameter space (where $t^{\prime}/t \in [0.1,1]$ and $n \in [0.5,1]$), the system systematically reveals symmetry-dependent magnetic correlation evolution results for the two-dimensional $t-t^{\prime}$ Hubbard model. 

We employ the constrained-path CPQMC method using a Hartree--Fock trial wave function~\cite{zhang1997, zhang1995}, which accurately captures quantum fluctuations and many-body correlations. By imposing a constrained-path approximation, CPQMC mitigates the fermion sign problem, enabling reliable simulations of the ground state properties away from half filling. We study the system with lattice size $L=16 \times 16$, which provides momentum space resolution to resolve key features in the Brillouin zone, including high-symmetry points $\Gamma=(0,0)$ and $X=(\pi,0)$. The hyper-parameters used in our CPQMC calculations are as follows: the number of walkers is $N_w=1000$; the number of blocks for relaxation is $N_{\mathrm{rel}}=10$; the number of blocks for growth estimate is $N_{\mathrm{gr}}=3$; the number of blocks after relaxation is $N_{\mathrm{meas}}=10$; the Trotter step size is $\Delta\tau=0.01$; \textcolor{black}{the corresponding projected length along the imaginary time axis prior to measurement is $\beta=32$;} and the growth-control energy estimate is $E_T=-50$. 

In this work, magnetic ordering tendencies are diagnosed from equal-time spin correlation, which can be evaluated from one-body Green's functions defined by fermionic creation and annihilation operators.
We introduce the (equal-time) one-body density matrix
\begin{equation}
G^{\sigma}_{ij}\equiv \left\langle \hat{c}_{i\sigma}\hat{c}^{\dagger}_{j\sigma}\right\rangle,
\label{eq:rho_def}
\end{equation}
where $\hat c^{\dagger}_{i\sigma}$ ($\hat c_{i\sigma}$) creates (annihilates) an electron with spin $\sigma$ on site $i$.
The local spin operator is
\begin{equation}
\hat S_i^z=\frac12\left(\hat n_{i\uparrow}-\hat n_{i\downarrow}\right),
\qquad
\hat n_{i\sigma}=\hat c^\dagger_{i\sigma}\hat c_{i\sigma}.
\label{eq:Sz_def}
\end{equation}
The equal-time spin correlation function is then
\begin{equation}
C^{zz}(i,j)\equiv \langle \hat S_i^z \hat S_j^z\rangle
=\frac14\Big\langle (\hat n_{i\uparrow}-\hat n_{i\downarrow})(\hat n_{j\uparrow}-\hat n_{j\downarrow})\Big\rangle.
\label{eq:Czz_def}
\end{equation}



\textcolor{black}{In ground-state CPQMC, the projected many-body state at projection step \(n\) is represented by an ensemble of weighted Slater-determinant walkers initialized as the trial wavefunction at the first step,$
|\Psi^{(n)}\rangle \propto \sum_k w_k^{(n)} |\phi_k^{(n)}\rangle$.
The ground state is obtained by the imaginary-time propagation using the second-order Trotter--Suzuki decomposition together with the Hubbard--Stratonovich transformation under the constrained-path condition of $\langle \Psi_T|\phi\rangle > 0$~\cite{zhang2019auxiliary}. } For a Slater-determinant walker $|\phi\rangle$, Wick's theorem reduces the required two-body density correlators to products of one-body Green's functions.
In particular, for $i\neq j$ one obtains

\begin{equation}
    \begin{aligned}
\langle \hat n_{i\sigma}\hat n_{j\sigma}\rangle_{\phi}
&= \langle \hat n_{i\sigma}\rangle_{\phi}\langle \hat n_{j\sigma}\rangle_{\phi} -\left|\langle \hat c_{i\sigma}\hat c^\dagger_{j\sigma}\rangle_{\phi}\right|^2\\
&= G^{\sigma}_{ii}(\phi)G^{\sigma}_{jj}(\phi)
-|G^{\sigma}_{ij}(\phi)|^2,
\label{eq:wick_same_spin}
    \end{aligned}
\end{equation}

\begin{equation}
    \begin{aligned}
\langle \hat n_{i\uparrow}\hat n_{j\downarrow}\rangle_{\phi}
=\langle \hat n_{i\uparrow}\rangle_{\phi}\langle \hat n_{j\downarrow}\rangle_{\phi}
=
G^{\uparrow}_{ii}(\phi)G^{\downarrow}_{jj}(\phi),
\label{eq:wick_diff_spin}
    \end{aligned}
\end{equation}
where $G^{\sigma}_{ij}(\phi)=\langle \hat c^\dagger_{i\sigma}\hat c_{j\sigma}\rangle_{\phi}$ is computed from the corresponding walker Green's function.
Substituting Eqs.~\eqref{eq:wick_same_spin}--\eqref{eq:wick_diff_spin} into Eq.~\eqref{eq:Czz_def} yields
\begin{equation}
    \begin{aligned}
C^{zz}(i,j)
=&
\frac14\Big(G^{\uparrow}_{ii}-G^{\downarrow}_{ii}\Big)\Big(G^{\uparrow}_{jj}-G^{\downarrow}_{jj}\Big)
- \\
&\frac14\Big(|G^{\uparrow}_{ij}|^2+|G^{\downarrow}_{ij}|^2\Big),
\qquad (i\neq j),
\label{eq:Czz_rho_compact}
    \end{aligned}
\end{equation}
and the CPQMC estimate is obtained by averaging $C^{zz}(i,j)$ over the walker ensemble (with the chosen estimator).

The spin structure factor is the Fourier transform of $C^{zz}(i,j)$,
\begin{align}
C^z(\mathbf k)=\frac{1}{N}\sum_{i,j}e^{i\mathbf k\cdot(\mathbf r_i-\mathbf r_j)}\,C^{zz}(i,j)
&=\frac{1}{N}\left\langle \hat S^z_{\mathbf k}\hat S^z_{-\mathbf k}\right\rangle,\\
S^z_{\mathbf k}=\sum_j e^{-i\mathbf k\cdot \mathbf r_j}\hat S^z_j.
\label{eq:Szk_from_Czz}
\end{align}

\textcolor{black}{
To connect with the real-space correlations shown below, we define the translationally averaged equal-time SDW correlation function
\begin{equation}
C_{\mathrm{SDW}}(\mathbf r)= \frac{1}{N}\sum_{j} C^{zz}(j+\mathbf r,j)
= \frac{1}{N}\sum_{j}\left\langle \hat S^z_{j+\mathbf r}\hat S^z_{j}\right\rangle .
\label{eq:CSDW_def}
\end{equation}
Here, $N$ is the total number of lattice sites, $j$ labels the reference site, and $\mathbf r$ denotes the displacement vector between the two sites.
}

Finally, we define the dominant magnetic ordering vector by the location of the maximum of $C^z(\mathbf k)$:
\begin{equation}
\mathbf Q=\arg\max_{\mathbf k} C^z(\mathbf k).
\label{eq:Qmax_def}
\end{equation}
This provides a unified criterion to classify the ordering vector throughout the phase diagram:
\begin{align}
\mathbf Q=&(\pi,\pi) \quad \Rightarrow \quad \text{N\'eel antiferromagnetism}, \nonumber\\
\mathbf Q=&(\pi,0)\ \text{or}\ (0,\pi) \quad \Rightarrow \quad \text{Stripe magnetism}, \nonumber\\
\mathbf Q=&(\pi,q)\ \text{or}\ (q,\pi),\ \ q\neq 0,\pi
\quad \Rightarrow \quad \text{Spiral magnetism}.
\label{eq:classification}
\end{align}
In this way, the magnetic order classification used in this work is directly rooted in Green's function based measurements: one-body Green's functions determine $C^{zz}(i,j)$ via Wick contractions, whose Fourier transform yields $C^z(\mathbf k)$ and hence the ordering vector $\mathbf Q$.

To distinguish between long-range magnetic order and short-range magnetic correlation, we employ a complementary analysis combining momentum space spectral features with real space correlation decay. In the momentum space, the static $C^z(\mathbf k)$ serves as the primary diagnostic indicator. A system possessing long-range order is characterized by sharp peak at specific $C^z(\mathbf k)$, with peak intensities remaining significantly enhanced even as system size increases. Conversely, short-range order manifests as broadened or diffuse weight centered around $\mathbf{Q}$. The finite width of these peaks in reciprocal space is inversely proportional to the magnetic correlation length in real space; thus, a diffuse peak signifies a finite short-range order, indicating that spin coherence is limited to local clusters. 

Through further analysis of the real space spin-spin correlation function $C_{\text{SDW}}(\mathbf{r})$, the differences between the two exhibit qualitative and distinct characteristics:
In the long-range sequence $C_{\text{SDW}}(\mathbf{r})$, the spatial oscillations maintain finite, nonzero amplitudes throughout the lattice, constrained only by the system boundaries in finite-size simulations. In contrast, the short-range correlation function exhibits exponential decay. Although local magnetic oscillations may persist over a few lattice spacings, the signal vanishes rapidly at large distances, confirming the absence of global phase coherence. By combining the sharpness of the $C^z(\mathbf k)$ peak with the asymptotic behavior of $C_{\text{SDW}}(\mathbf{r})$, the nature of magnetic order can be reliably identified.

\subsection{B. Van Hove Singularities}
\label{VHS DISTRIBUTION}


\textcolor{black}{In this section, we analyse single-particle dispersion under the non-interacting limit U=0 to locate the VHS and and evaluate how Fermi surface geometry and VHS affect magnetic correlation.}  We evaluate the partial derivatives of the band dispersion $\varepsilon (\mathbf{k})$ with respect to the momentum components, and determine the stationary points at which these derivatives vanish.
\begin{equation}
    \begin{aligned}
&\frac{\partial \varepsilon }{\partial k_{x} } = 2t\sin k_{x} \pm 4t^{'}\cos k_{x}\sin k_{y}, \\
&\frac{\partial \varepsilon }{\partial k_{y} } = 2t\sin k_{y} \pm 4t^{'}\cos k_{y}\sin k_{x}.
    \end{aligned}
\end{equation} 
the VHS represent features of the band structure that occur at stationary points of the dispersion $\nabla_{\mathbf{k}}\varepsilon_{\mathbf{k}\sigma}=0$~\cite{van1953occurrence}. \textcolor{black}{In two dimensions, VHS produces a strong enhancement of DOS, thereby amplifying the impact of electron correlations: when the Fermi level approaches a VHS, even moderate (or weak) onsite interaction $U$ can trigger pronounced many-body instabilities and drive spontaneous symmetry breaking. Related VHS-driven competition among magnetic orders, including altermagnetic phases, has also been discussed recently in multi-orbital settings~\cite{LuHu2026}. Moreover, the distribution and evolution of VHS points (e.g., a shift from the $M=(\pi,0)$ and $(0,\pi)$ point into the interior of the Brillouin zone) reshapes the Fermi surface geometry and the dominant nesting vectors, providing a natural momentum-space filter that selects which ordering wave vector is most strongly enhanced. This VHS-guided kinematic selection offers a concise rationale for the observed evolution of the dominant magnetic correlations, such as the crossover from N\'eel-type $(\pi,\pi)$ tendencies to spiral $(\pi,q)$ textures upon tuning $t'$ and filling.} 

VHS generates a logarithmic enhancement of DOS, which strongly amplifies low-energy phase space for both particle-hole and particle-particle processes, 
\textcolor{black}{but the nature of the DOS singularity is controlled by the local band curvature encoded in the Hessian
$\mathcal H_{ij}(\mathbf k)=\partial^2\epsilon/\partial k_i\partial k_j$ \cite{yuan2019magic}.
For an ordinary two-dimensional saddle point, $\det(\mathcal H)\neq 0$ and the quadratic expansion around $\mathbf k^\ast$ is non-degenerate,
\begin{equation}
\epsilon(\mathbf k)\approx \epsilon^\ast + a\,\delta k_x^2 - b\,\delta k_y^2,\qquad a,b>0,
\end{equation}
which yields the characteristic logarithmic enhancement of the DOS,
$N(E)\propto \ln\!\big(\Lambda/|E-\epsilon^\ast|\big)$ (here $\Lambda$ is an energy cutoff that reflects the limited validity of the local quadratic expansion near the stationary point).
In our band structure this corresponds to the noncritical regimes away from the Lifshitz point (e.g., $t'<0.5$ and $t'>0.5$), where the saddle remains quadratic with finite principal curvatures.}
\textcolor{black}{At the critical value $t'=0.5$, the system undergoes a Lifshitz transition in which the saddle becomes degenerate:
not only does $\nabla_{\mathbf k}\epsilon=0$ at the stationary point, but one eigenvalue of $\mathcal H$ approaches zero, so the quadratic term is insufficient along the corresponding direction and higher-order terms dominate \cite{yuan2020classification}. A minimal local form is then
\begin{equation}
\epsilon(\mathbf k)\approx \epsilon^\ast + c\,\delta k_x^{4}- b\,\delta k_y^{2},
\end{equation}
leading to a stronger, higher-order VHS(HO-VHS) with a power-law singularity,
$N(E)\propto |E-\epsilon^\ast|^{-1/4}$ \cite{yuan2019magic, shtyk2017electrons, isobe2019supermetal}.
This enhanced DOS reflects an unusually flat dispersion and implies a substantial amplification of interaction-driven instabilities near the corresponding filling.}
\textcolor{black}{Upon increasing $t'$ beyond $0.5$, the HO-VHS generically bifurcates into multiple ordinary quadratic saddles. Locally, the Hessian becomes non-degenerate again at each new saddle, restoring the logarithmic form of the singularity. Importantly, although the exponent reverts to the ordinary VHS behavior, the multiplicity of saddle points increases, which can maintain a large overall DOS enhancement and thus sustain strong competition among magnetic ordering tendencies in the interacting CPQMC results.}
We computed the VHS structure for $t' \in [0.1,0.9]$ and summarize the complete set of results in Appendix~B. In the main text, we focus on three representative values, $t'=0.1$, $0.5$, and $0.9$, which capture the key trends across this range. 

\textcolor{black}{In the regime of small $t'$  (e.g., $t'=0.1$), the VHS of the noninteracting dispersion are located at the time-reversal invariant momenta $(\pi,0)$ and $(0,\pi)$. These points are ordinary two-dimensional saddle points: $\nabla_{\mathbf k}\epsilon(\mathbf k^\ast)=0$ and the Hessian $\mathcal H_{ij}(\mathbf k^\ast)$ is non-degenerate, $\det\mathcal H(\mathbf k^\ast)\neq 0$, with two principal curvatures of opposite signs. As a consequence, the DOS exhibits the characteristic logarithmic enhancement,
$N(\varepsilon)\propto \ln\!\big(\Lambda/|\varepsilon-\varepsilon_{\rm VHS}|\big)$.
In this noninteracting setting, the corresponding Fermi surface geometry implies an enhanced particle--hole phase space near a nesting $(\pi,\pi)$, which provides a kinematic tendency toward N\'eel-type correlations \cite{hlubina1997ferromagnetism}.}

A transition occurs at the critical value $t' = 0.5$. Here, the VHS shift to the $\Gamma$ point $(0,0)$ and the $K$ point $(\pi, \pi)$. Crucially, the curvature along the node direction ($k_x = \pm k_y$) approaches zero, leading to a significant enhancement in curvature divergence. This flat-band dispersion leads to a HO-VHS, characterized by a stronger, power-law divergence in the DOS, $N(\varepsilon) \sim |\varepsilon|^{-1/4}$ \cite{yuan2019magic, shtyk2017electrons, isobe2019supermetal}. According to the Stoner criterion, this divergent DOS significantly amplifies the influence of electron correlation, driving the system towards a complex non-collinear ordering, even at moderate interaction strengths \cite{schulz1987superconductivity}.

\begin{figure}[b!]
    \centering
    \includegraphics[width=0.5\textwidth]{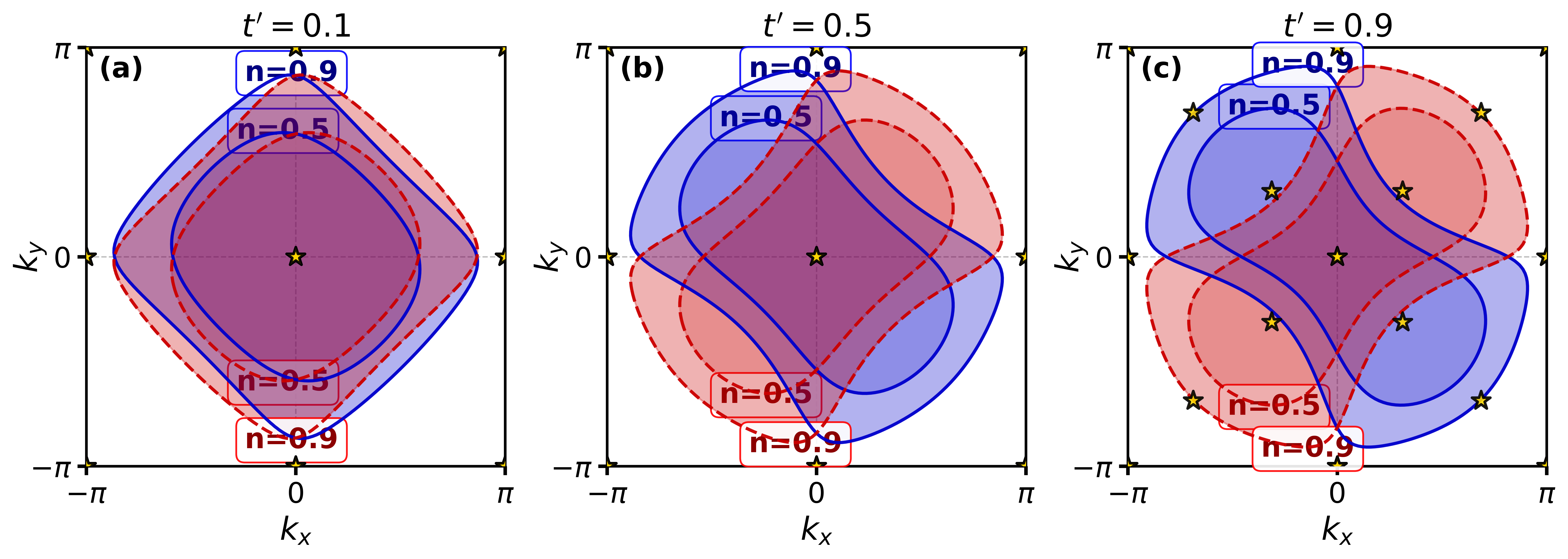}
    \caption{
The evolution of VHS and Fermi surfaces with varying NNN hopping $t'$. The panels correspond to (a) $t'=0.1$, (b) $t'=0.5$, and (c) $t'=0.9$. For each $t'$, solid (dashed) contours denote the Fermi surfaces of the upper (lower) band at fillings $n = 0.5$ and $n = 0.9$, while the shaded regions indicate the corresponding filled Fermi surface. Stars mark the VHS. As $t'$ increases, the VHS manifests not only at high symmetry points and the Fermi surfaces become strongly distorted, degrading $ (\pi,\pi)$ nesting and thereby favoring $ ( \pi, q)$ or $ ( q, \pi)$ spiral tendencies and, at larger $t'$, stripe collinear configurations $ (\pi,0)$ or $ (0,\pi)$.}
    \label{fig:my_label8}
\end{figure}

\textcolor{black}{For large $t'$ (e.g., $t'=0.9$), the higher-order singularities at the high-symmetry points generically bifurcate into multiple ordinary saddle points and migrate to incommensurate positions in the Brillouin zone.Crucially, the DOS singularity's regression to the standard logarithmic form stems from the non-degeneracy of the local quadratic expansion at each new saddle point: the Hessian matrix $\mathcal H_{ij}(\mathbf k^\ast)$ possesses two non-zero eigenvalues of opposite sign. Consequently, $N(\varepsilon)$ exhibits standard two-dimensional VHS behaviour:
$N(\varepsilon)\propto \ln\!\big(\Lambda/|\varepsilon-\varepsilon_{\rm VHS}|\big)$
Moreover, due to increased saddle point multiplicity, the overall amplification effect may remain substantial.
At the same time, the $t'$-driven reshaping of the Fermi surface weakens the near-$(\pi,\pi)$ particle--hole phase space and shifts the dominant nesting away from $(\pi,\pi)$. This naturally disfavors N\'eel-type $(\pi,\pi)$ tendencies and promotes incommensurate spiral textures or collinear stripe correlations.}

As shown in Fig.~\ref{fig:my_label8}, in the present spin splitting NNN Hubbard model, the location and energetic relevance of the VHS are strongly controlled by the NNN hopping amplitude $t'$.  In particular, increasing $t'$ continuously shifts the saddle-point energy relative to the Fermi level and simultaneously deforms the Fermi surface, thereby modifying the dominant low-energy scattering channels.  Our results reveal a qualitative boundary around $t'\!\approx\!0.5$: for $t'\lesssim 0.5$ the VHS remains `close' to the conventional square-lattice saddle-point sector and the Fermi surface retains substantial diagonal connectivity, whereas for $t'\gtrsim 0.5$ the band reconstruction driven by spin-dependent NNN hopping qualitatively reshapes the Fermi surface into strongly anisotropic pocket-like structures, indicating that the system has crossed into a distinct kinematic regime. In this sense, $t'\!\approx\!0.5$ acts as an effective demarcation line separating two different VHS-controlled Fermi surface and, correspondingly, two different magnetic-selection mechanisms.
\begin{figure}[H]
    \centering
    \includegraphics[width=0.5\textwidth]{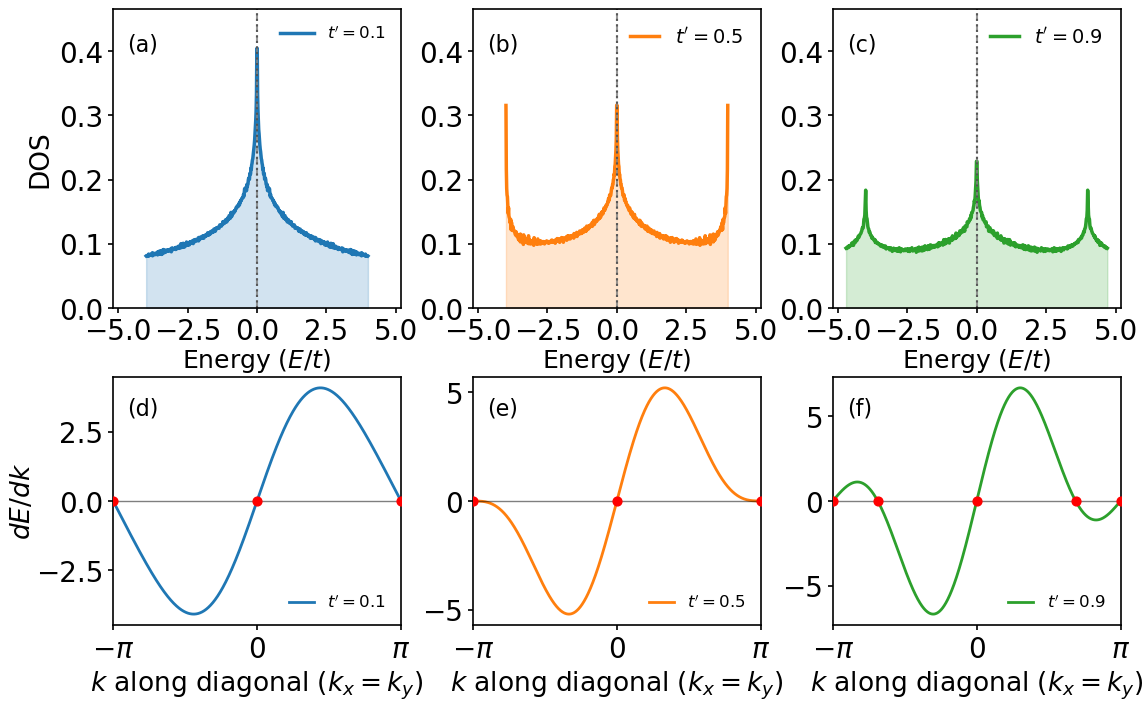}
    \caption{
The evolution of the electronic DOS as a function of the NNN hopping amplitude $t'$.
The panels (a)–(c) display the calculated total DOS for $t' = 0.1$, $t' = 0.5$, and $t' = 0.9$, respectively. The energy $E$ is normalized by the NN hopping $t$. Vertical dashed lines mark the energy $E=0$, typically corresponding to the Fermi level at half-filling. The prominent peaks in the DOS spectra identify the positions of VHS, which arise from saddle points in the momentum space dispersion relation.
The panels (d)–(f) depict the corresponding  $\nabla_{\mathbf{k}}\varepsilon_{\mathbf{k}\sigma}$ along the diagonal momentum cut. The red circles mark the critical points where the $\nabla_{\mathbf{k}}\varepsilon_{\mathbf{k}\sigma}$ vanishes, directly corresponding to the VHS locations. }
    \label{fig:my_label9}
\end{figure}
To corroborate the evolution of the electronic structure, we calculate the total DOS and the $\nabla_{\mathbf{k}}\varepsilon_{\mathbf{k}\sigma}$ along the nodal diagonal ($k_x=k_y$), as shown in Fig.~\ref{fig:my_label9}.
For $t'=0.1$, the DOS exhibits a conventional logarithmic divergence characteristic of a standard saddle point Fig.~\ref{fig:my_label9}(a). The group velocity along the diagonal vanishes linearly only at the high-symmetry points $\Gamma$ ($k=0$) and $M'$ ($k=\pm\pi$), consistent with a quadratic band dispersion in Fig.~\ref{fig:my_label9}(d). A critical transition occurs at $t'=0.5$. \textcolor{black}{Here, at the centre of the region shown in Fig.~\ref{fig:my_label9}(e), $\nabla_{\mathbf{k}}\varepsilon_{\mathbf{k}\sigma}$ approaches zero in the tangential direction, indicating the presence of a stationary point where dispersion along the diagonal exhibits pronounced flattening characteristics (i.e., the corresponding second-order derivative diminishes, and higher-order terms commence to exert influence).}. This flattening of the dispersion confirms the emergence of a HO-VHS, where the dispersion is dominated by quartic terms ($\sim k^4$). Consequently, the DOS singularity becomes sharper, reflecting the divergence of the effective mass in Fig.~\ref{fig:my_label9}(b). In the strong spin splitting ($t'=0.9$), the single VHS at $\Gamma$ bifurcates. This splitting is explicitly captured by the group velocity profile in Fig.~\ref{fig:my_label9}(f), where new nodes (marked by red circles) emerge at momenta away from $k=0$. These nodes correspond directly to the split peaks observed in the DOS in Fig.~\ref{fig:my_label9}(c), signifying the migration of VHS into the Brillouin zone interior.

\section{\uppercase\expandafter{\romannumeral 3}. RESULTS}
\label{Result}
\subsection{A. Magnetic phase diagram}
\label{Phase diagram}
\begin{figure}[H]
    \centering
    \includegraphics[width=0.5\textwidth]{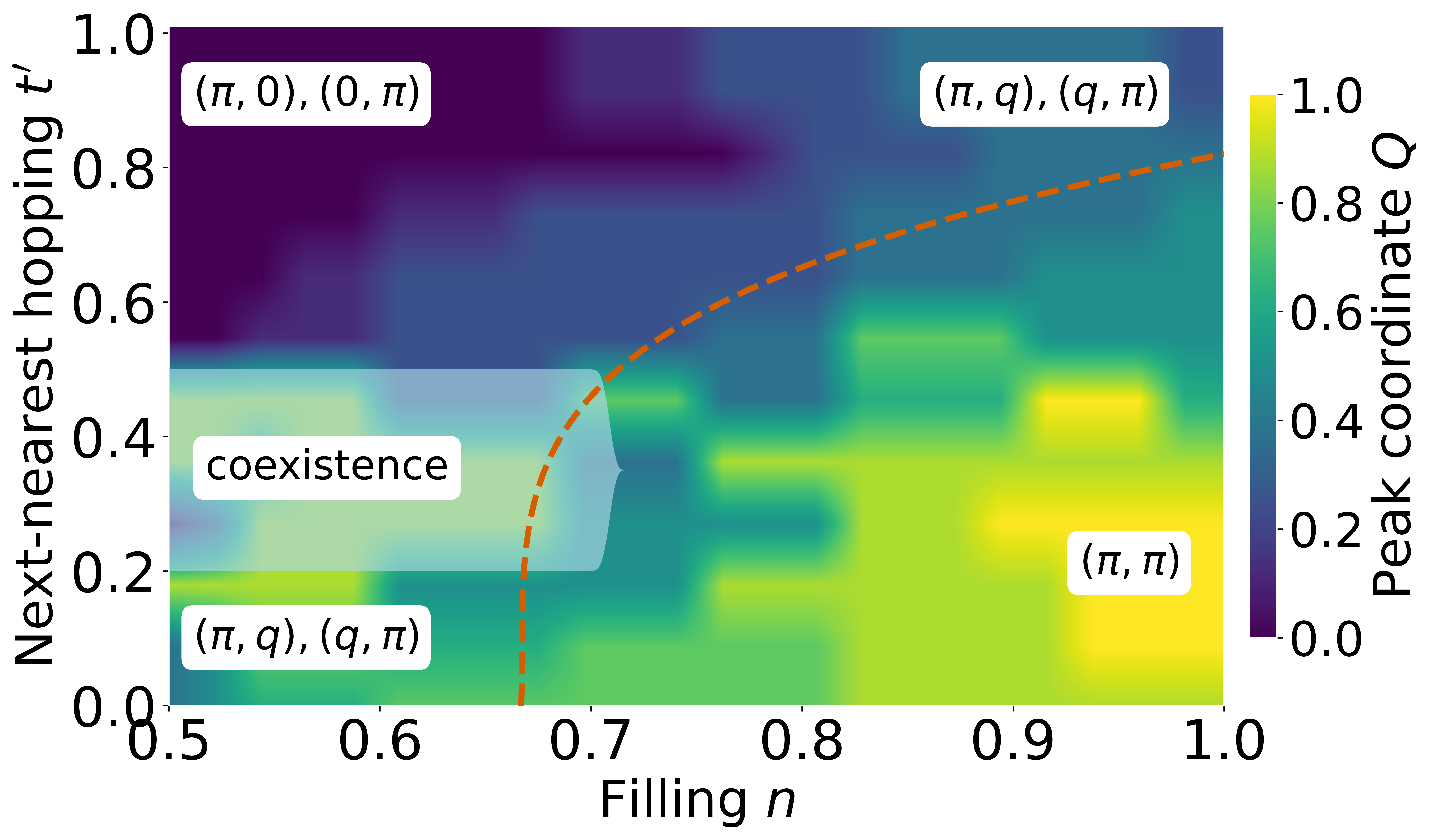}
    \caption{
The ordering vector distribution of dominant $C^z(\mathbf k)$ peak locations in the $(n,t')$ plane for an $L=16$ lattice.
The color scale shows the $C^z(\mathbf k)$ peak coordinate $Q$ at which $|C^z(\mathbf k)|$ is maximal:
$Q=1$ corresponds to N\'eel order at $(\pi,\pi)$, $Q=0$ to collinear stripe correlation at $(\pi,0)$ or $(0,\pi)$, and $0<Q<1$ indicates spiral correlation, e.g. $(\pi,q)$ or $(q,\pi)$.
The shaded region labeled coexistence regimes where several $C^z(\mathbf k)$ peak types have comparable maximal weight , indicating strong competition among magnetism tendencies. Ordered states with different magnetic correlation are distinguished by a red line: regions to the right of the red line represent long-range order, while those to the left indicate short-range order.
Overall, the diagram illustrates the evolution from N\'eel/stripe regimes to spiral correlation behavior as $t'$ varies.
}
    \label{fig:my_label3}
\end{figure}

\textcolor{black}{We now present the CPQMC results for the interacting case at $U=4$. The complementary noninteracting ($U=0$) calculations, together with their analysis and discussion, are deferred to the Appendix C.} The distribution of $C^z(\mathbf k)$ peaks as functions of the electron filling \(n\) and NNN hopping \(t'\) is shown in Fig.~\ref{fig:my_label3}. The figure clearly demonstrates the transition between different types of magnetic correlation, with antiferromagnetic order, spiral correlation, and collinear stripe magnetic correlation exhibiting consistent patterns across various lattice sizes. In the low-doping regime, the system stabilizes into N\'eel antiferromagnetic order, which is characterized by a pronounced $C^z(\mathbf k)$ response at the \( (\pi, \pi)\) point in momentum space. As \(t'\) increases, the system's magnetism undergoes a gradual transition towards a non-collinear magnetic state, with peaks shifting to non-collinear $C^z(\mathbf k)$, particularly at \( (\pi, q)\) or \( (q, \pi)\). 

We examined the magnetic phase diagram at different lattice sizes and found that the overall magnetic behavior exhibits robustness to variations in system size under the same analytical framework. Specifically, phase boundaries and qualitative ordering trends remain consistent across different parameter ranges, with relevant results summarized in Appendix A. These observations point to a progressive transition from N\'eel antiferromagnetic order to stripe and spiral states as the doping concentration \(n\) and the NNN hopping \(t'\) are varied. The NNN hopping term \(t'\) introduces a momentum-dependent spin splitting, causing distinct spin branches to have different energies in momentum space. This could result in the non-collinearity of the spin alignment. The phase boundary distinguishing long-range magnetic order from short-range correlation is delineated in the Fig.~\ref{fig:my_label3}. \textcolor{black}{The red dashed boundary in Fig.~\ref{fig:my_label3} is extracted from the peak-sharpness metric $R_p$, defined as $R_p=(C_{\max}-\langle C_{\mathrm{near}}\rangle)/C_{\max}$, where $C_{\max}$ is peak value of $|C^{z}(\mathbf{k})|$ and $\langle C_{\mathrm{near}}\rangle$ is the average over the nearest momentum points around the dominant peak. Because the momentum-space correlation function is the Fourier transform of the real-space correlator, a sharper peak in momentum space generally reflects correlations that remain coherent over a longer spatial range. Therefore, $R_p$ provides a measure for the spatial coherence of magnetic correlations. The boundary is obtained by fitting the resulting crossover points in the $(n,t')$ plane [see Fig.~\ref{fig:appendix_3}(b)].} The long-range phase is predominantly associated with the ordering vector $\mathbf{Q}=(\pi, \pi)$, $(\pi, q)$ or $(q, \pi)$, and is characterized by the emergence of sharp, well-defined peaks—as opposed to diffuse features—in the SDW momentum space spectrum. Notably, this long-range order demonstrates robustness against finite size effects, a property that will be elaborated upon in the subsequent sections.

As \(t'\) increases, the system transitions from a collinear magnetic ordered state to a stripe or spiral configuration, with the corresponding ordering vector shifting from \( (\pi, \pi)\) to \( (\pi, q)\), thus demonstrating a continuous change in the system's magnetic phase driven by the interplay between electronic interactions and the hopping parameters. For fillings $n\simeq 0.5$ and small NNN hopping $t'$, the $C^z(\mathbf k)$ develops pronounced peaks at both $\mathbf{Q}= (\pi,\pi)$ and $ (\pi,0)$ or $ (0,\pi)$, this area is referred to as the coexistence regime. \textcolor{black}{This double-peak structure indicates strong competition between N\'eel and stripe SDW tendencies. It means that the structure factor has two comparable maxima at $(\pi,\pi)$ and $(\pi,0)$ or $(0,\pi)$, rather than a single dominant ordering vector.}  The quasi-square Fermi contour retains strong particle--hole scattering at $\mathbf{Q}=(\pi,\pi)$, thereby stabilizing N\'eel antiferromagnetism. Upon doping away from half filling, the contour develops elongated axial features that enhance low-energy scattering at $\mathbf{Q}=(\pi,0)$ or $(0,\pi)$, which in turn promotes collinear stripe correlation.

At high filling (\(n \sim 1\)) and low NNN hopping strength (\(t' \sim 0\)), the magnetism of the system is primarily driven by electron-electron interactions. In this regime, as shown in Fig.~\ref{fig:my_label3}, for \(n = 0.953\) and \(t' = 0.1\), perfect Fermi surface nesting results in the antiferromagnetic ordering vector \(\mathbf{Q} = (\pi, \pi)\) being the most energetically favourable configuration, stabilising N\'eel antiferromagnetic order. As the doping concentration \(n\) decreases and \(t'\) increases, the NNN hopping \(t'\) introduces momentum-dependent spin splitting, where distinct spin branches exhibit different energy levels in momentum space. For instance, when \(n = 0.789\) and \(t' = 0.5\) and \(n = 0.945\) with \(t' = 0.5\), the antiferromagnetic order stability is disrupted, with peaks appearing at \( (\pi, q)\) or \( (q, \pi)\), indicating a transition towards a spiral magnetism. When \(n\) is small and \(t' \sim 0.4\), the system cannot maintain its initial antiferromagnetic N\'eel state, increasingly tending towards non-collinear spin configurations (such as the spiral state) or collinear stripe state. The ordering vector shifts from the collinear point \( (\pi, \pi)\) towards non-collinear points (e.g., \( (\pi, q)\)) or the collinear point \( (\pi, 0)\), when \(n=0.633\) and \(t' \le 0.5\), N\'eel antiferromagnetic order coexists with collinear stripe magnetic correlation.  Upon further decreasing \(n\) and increasing \(t'\), the system response evolves into a collinear state, where the spin distribution corresponding to collinear stripe magnetic correlation becomes uniform. As \(n\) further decreases and \(t' \sim 1\), the spin distribution in real space approaches uniformity, with the dominant magnetism being fully transformed into collinear stripe correlations, the ordering vector transitions to either \( (\pi, 0)\) or \( (0, \pi)\). 

\subsection{B. From N\'eel \((\pi, \pi)\) antiferromagnetism to spiral \((\pi, q)\) magnetism}
\label{AFM}
\begin{figure}[b!]
    \centering
    \includegraphics[width=0.5\textwidth]{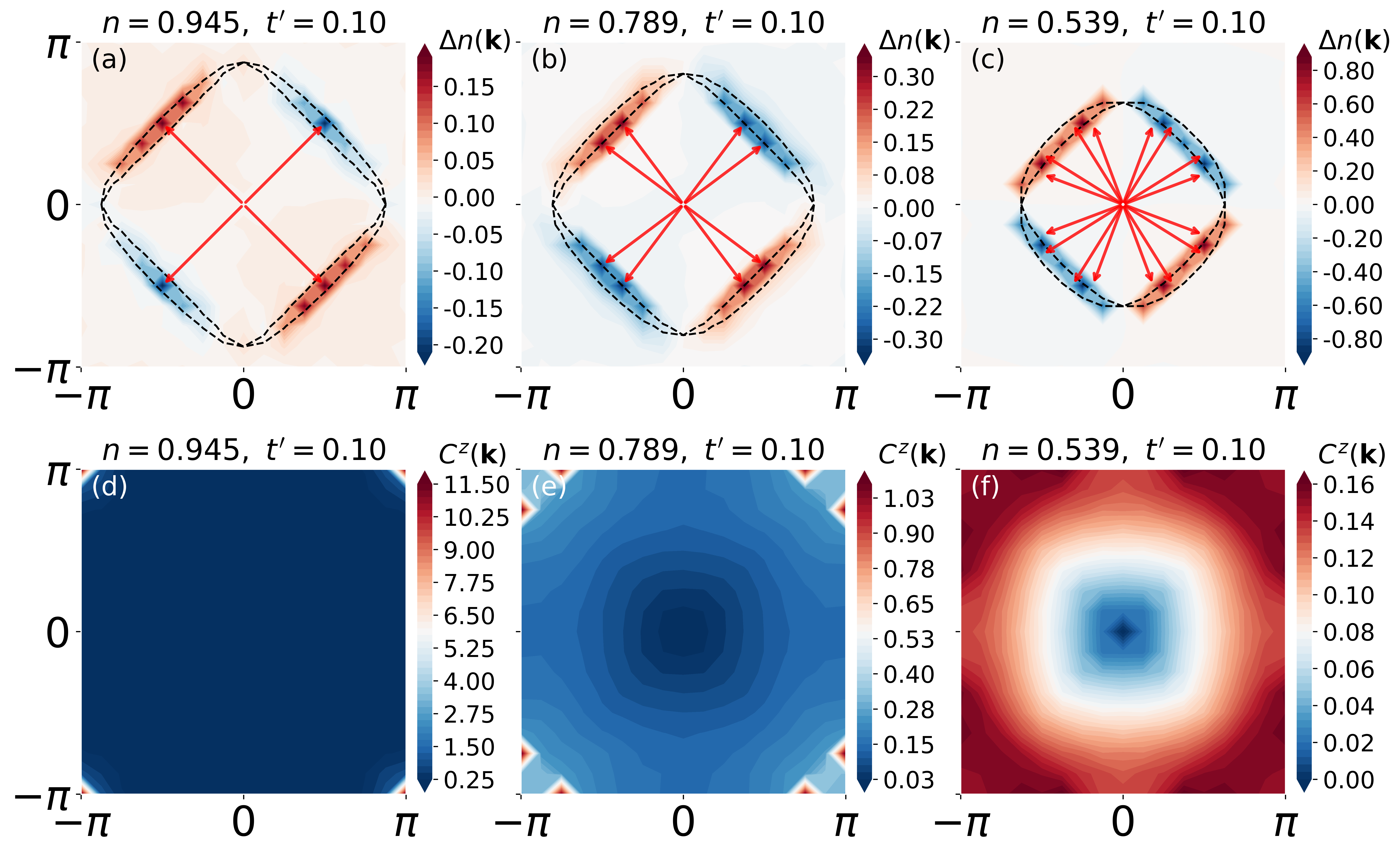}
    \caption{
The momentum-resolved spin imbalance $\Delta n(\mathbf{k})$ and magnetic correlation at weak spin anisotropy for fixed $t'=0.10$. (a)–(c) $\Delta n(\mathbf{k})$ for fillings $n=0.945$, $0.789$, and $0.539$, respectively, overlaid with the noninteracting Fermi surfaces (dashed) and the dominant magnetic wave vector (red arrows).
(d)–(f) Corresponding $C^z(\mathbf k)$ for the same parameters.
Near half filling, $C^z(\mathbf k)$ is peaked at $(\pi,\pi)$, while upon decreasing $n$ the dominant response shifts toward ordering vector of the form $(\pi,q)$ or $(q,\pi)$, accompanied by a splitting of the $C^z(\mathbf k)$ peak structure.
}
    \label{fig:my_label4}
\end{figure}

We study how the non-interacting Fermi surface, the corresponding total spin polarization $\Delta n^{\mathrm{tot}} (\mathbf{k})$, the magnetic vector textures, and the $C^z (\mathbf{k})$ change with a small spin–anisotropic NNN hopping $t'=0.1$ and varying doping, as shown in Fig~\ref{fig:my_label4}. The noninteracting band dispersion [Eq.~(2)] implies that, as $n \to 1$ and in the limit $t' \to 0$, one approximately recovers the nesting relation
\begin{equation}
\varepsilon_{\mathbf{k}+\mathbf{Q},\sigma} \simeq -\varepsilon_{\mathbf{k}\sigma},
\qquad \mathbf{Q}= (\pi,\pi),
\end{equation}
each spin–resolved Fermi surface is almost perfectly nested by the ordering vector $\mathbf{Q}= (\pi,\pi)$. For the small but finite value $t'=0.1$ the spin splitting is weak, and this nesting condition is only mildly perturbed. At densities close to half filling ($n\approx 1$), the Fermi level lies very near the VHS energy of the band: the Fermi surface passes through the $M$ points and forms an almost square contour rotated by $45^\circ$ with respect to the Brillouin-zone axes, intersecting the VHS exactly. This geometry produces a large density of states at the Fermi level and strong nesting, since the ordering vector $\mathbf{Q}= (\pi,\pi)$ maps one side of the Fermi surface onto the opposite side. Consequently, the magnetic vector points towards $ (\pi,\pi)$ and exhibits a pronounced peak at the $C^z(\mathbf k)$, favouring the formation of N\'eel antiferromagnetic order. This behavior is clearly visible in Fig.~\ref{fig:my_label4}(a), at $n\simeq 0.95$ the $C^z (\mathbf{k})$ is sharply peaked at $ (\pi,\pi)$ and the magnetic vector field reflects a $ (\pi,\pi)$ modulation.

Upon reducing the filling, the energy band edges move away from the VHS, the Fermi surface shrinks toward the Brillouin-zone center, and its shape is progressively distorted by the $t'$ term, as shown in Fig.~\ref{fig:my_label4}. As a result, the diagonal segments that were well nested by $ (\pi,\pi)$ become more curved and the corresponding nesting rapidly deteriorates, while the Fermi surfaces cannot be directly mapped via $ ( \pi, \pi ) $, but are more effectively connected through ordering vector of the form $\mathbf{Q} \approx (\pi, q)$ or $ ( q, \pi) $. Geometrically, the Fermi surface evolves from a square structure towards a more axially elongated shape, whilst the momentum $\mathbf{Q}$ at which $C^z(\mathbf k)$ attains its maximum shifts from the conjugate point $ (\pi,\pi)$ towards the non-conjugate axial ordering vector $ ( \pi, q)$ or $ ( q, \pi)$, as illustrated in Fig.~\ref{fig:my_label4}. This evolution is reflected in the momentum–resolved spin imbalance $\Delta n^{\sigma} (\mathbf{k})$ , where the dominant scattering “hot spots” move from the diagonal regions to the antinodal patches, and in the $C^z(\mathbf k)$ , where the main peak gradually splits and drifts away from $ (\pi,\pi)$ as $n$ decreases. Together, these observations demonstrate that the optimal fermi surface nesting vector of the spin sector changes continuously from a $\mathbf{Q}= (\pi,\pi)$ near half filling to $\mathbf{Q}= ( \pi, q)$ at lower densities, driving the transition from antiferromagnetism to spiral magnetism.
\begin{figure}[b!]
    \centering
    \includegraphics[width=0.5\textwidth]{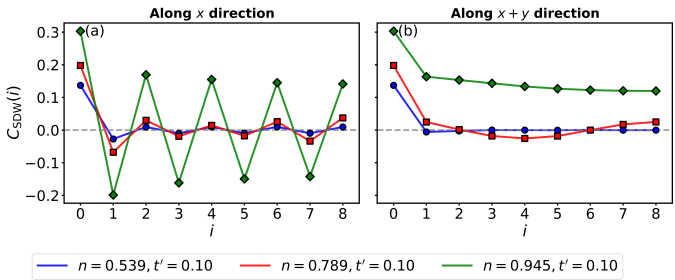}
    \caption{
The real space magnetic correlation function $C_{\mathrm{SDW}}(i)$ at fixed $t'=0.1$ for three representative fillings ($n=0.945$, $0.789$, and $0.539$).
(a) correlation along the $x$ direction and (b) along the $x+y$ direction.
Near half filling ($n=0.945$), $C_{\mathrm{SDW}}(i)$ exhibits a robust staggered sign structure with sizeable amplitude over the accessible distances, consistent with N\'eel antiferromagnetic order.
Upon reducing the filling (e.g., $n\approx 0.5$), the staggered component decays rapidly, indicating that magnetic correlation become short-ranged away from half filling.}
    \label{fig:my_label4b}
\end{figure}

Simultaneously, to visualise the magnetic correlation pattern in real space, we perform an inverse Fourier transform of the longitudinal $C^z(\mathbf k)$ obtained in momentum space.  
The result of the inverse Fourier transform reflects the property of the spontaneous symmetry-broken correlation function $C^z(\mathbf k) = C^z(-\mathbf k)$, yielding the spin correlation function in real space.
Fig.~\ref{fig:my_label4b} displays the resulting magnetic correlation pattern in real space for representative values of $t'$ and filling. For $t'=0.1$ and fillings close to half filling, the real space magnetic correlation exhibits a robust staggered pattern whose amplitude remains sizeable over the entire lattice, signalling long-range N\'eel antiferromagnetic order. In contrast, when the system is tuned far away from half filling (e.g., around $n\simeq 0.5$), the magnetic correlation decay rapidly with distance and the staggered pattern becomes only partially developed, indicating that the magnetism sequence has been simplified to short-range correlation. For $n=0.789$, a finite long-range correlation persists; although the magnitude is suppressed, the signal remains robust without decaying as the spatial separation increases.
\begin{figure}[b!]
    \centering
    \includegraphics[width=0.5\textwidth]{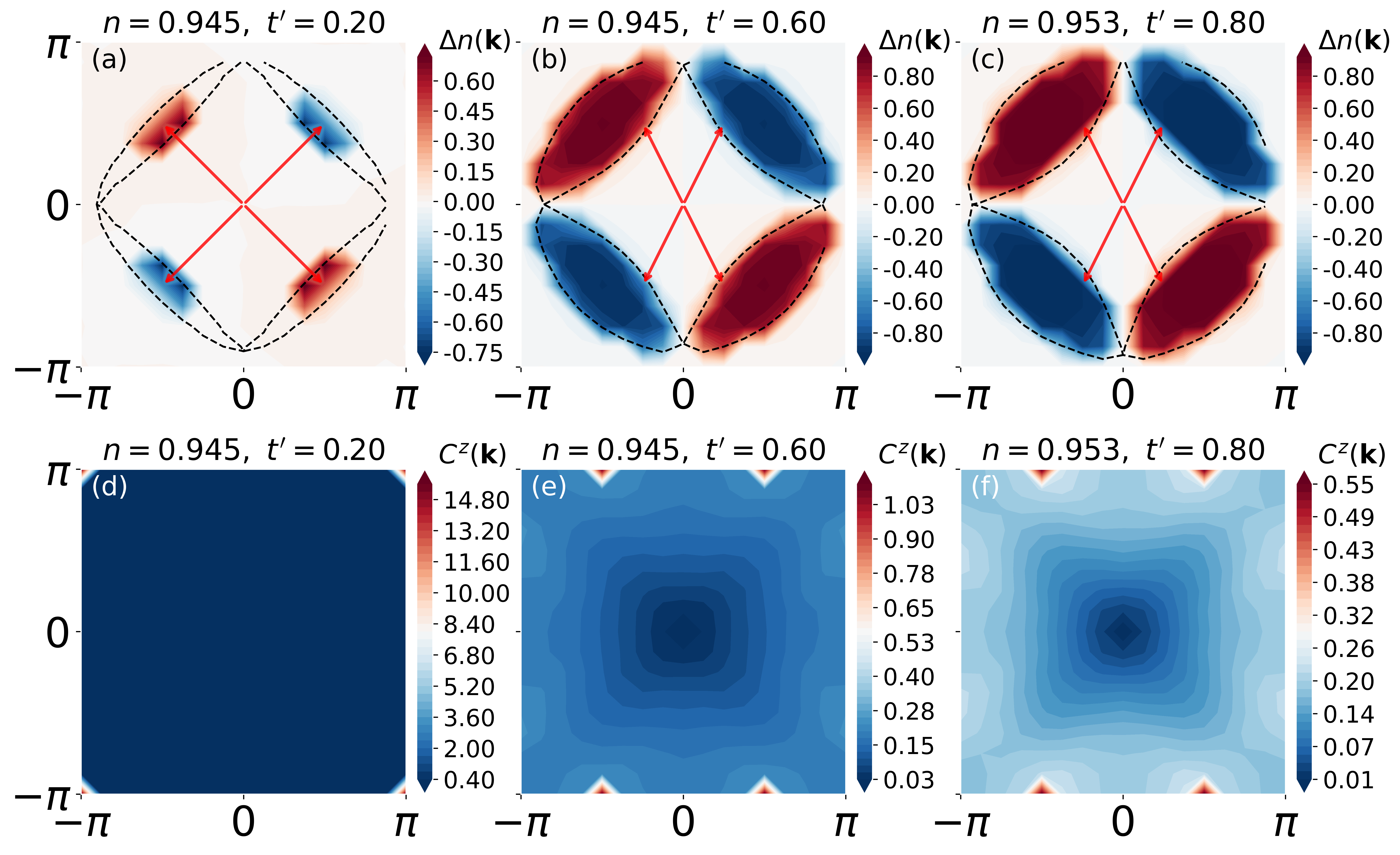}
    \caption{
The momentum-resolved spin imbalance and magnetism response for $n \sim  1$. 
(a)–(c) Spin imbalance $\Delta n(\mathbf{k})$ for the parameter sets $(n,t')=(0.945,0.20)$, $(0.945,0.60)$, and $(0.953,0.80)$, respectively, overlaid with the corresponding noninteracting Fermi surfaces (dashed). 
(d)–(f) Corresponding $C^z(\mathbf k)$ for the same parameters as in (a)–(c). 
Compared with the weak-$t'$ regime, the $C^z(\mathbf k)$ spectral weight becomes broadened and is redistributed away from a sharp $(\pi,\pi)$ response, consistent with the loss of near-perfect nesting and the emergence of frustrated spiral magnetic correlation tendencies.
}
    \label{fig:my_label6}
\end{figure}

For larger values of spin anisotropy near half-filling, the system manifests the axial ordering vector of the form $\mathbf Q= (\pi,q)$ or $ (q,\pi)$. This can be understood within a Fermi surface nesting picture. As shown in Fig.~\ref{fig:my_label6}, once the spin–dependent NNN hopping enters the intermediate regime $t'\gtrsim 0.5$, $\Delta n^\sigma(\mathbf{k})$ exhibits four well-defined pocket-like features, one in each quadrant of the Brillouin zone. This pocket reconstruction implies that the low-energy particle--hole phase space becomes highly anisotropic: rather than being dominated by the diagonal translation $\mathbf{Q}=(\pi,\pi)$ (characteristic of the weak-$t'$ regime) or by the purely axial stripe ordering vector $(\pi,0)/(0,\pi)$ (favored at very large $t'$), the most efficient scattering processes now connect distinct pocket segments with comparable curvature and orientation through a continuous set of ordering wave vectors of the form\[\mathbf{Q}\simeq (\pi,q)\quad \text{or}\quad (q,\pi),\qquad 0<q<\pi.\]
\textcolor{black}{we also note that in the larger-$t'$ regime the interacting system does not always treat the two symmetry-related orientations $(\pi,q)$ and $(q,\pi)$ equivalently in the momentum-resolved maps. We interpret this as a genuine interaction-driven breaking of the fourfold ($C_4$) rotational symmetry in the correlated state, rather than an artifact of the CPQMC method.} In a weak-coupling description, this is precisely the regime in which the susceptibility develops maxima at spiral momenta: the dominant contribution to susceptibility is generated by nearly parallel (low-curvature) Fermi Surface segments, and the characteristic mismatch between diagonal and axial nesting renders the susceptibility peak shifted away from the high-symmetry points. 

This mechanism is directly reflected in $C^z(\mathbf k)$. In Fig.~\ref{fig:my_label6}, the spectral weight of $C^z(\mathbf k)$ broadens and is redistributed away from a sharp $(\pi,\pi)$ peak, with the dominant weight shifted toward spiral ordering wave vectors of the form $\mathbf{Q}\simeq(\pi,q)$ or $(q,\pi)$. This diffuse peak structure indicates that a range of nearby particle--hole scattering processes contribute comparably, consistent with a susceptibility landscape that is relatively shallow in momentum space around the optimal $\mathbf{Q}$.
\begin{figure}[b!]
    \centering
    \includegraphics[width=0.5\textwidth]{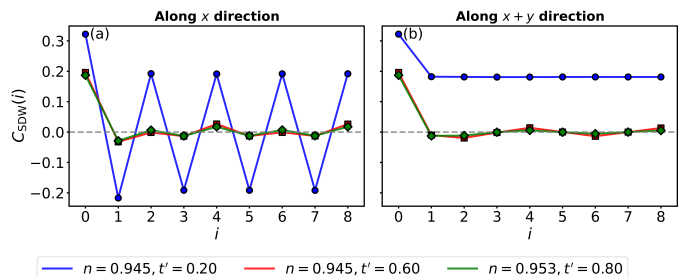}
    \caption{
The real space magnetic correlation function $C_{\mathrm{SDW}}(i)$ obtained from the inverse Fourier transform of $C^z(\mathbf k)$ for the same three fillings shown in Fig.~\ref{fig:my_label6}. 
(a) correlation evaluated along the $x$ direction and (b) along the $x+y$ direction. 
In both cases, $C_{\mathrm{SDW}}(i)$ exhibits a dominant on-site component followed by rapidly decaying, weakly oscillatory correlation, indicating long-range magnetic order in the frustrated regime.}
    \label{fig:my_label6b}
\end{figure}

The spatial correlation function $C_{\mathrm{SDW}}(i)$ exhibits oscillatory behavior with relatively small amplitudes as shown in Fig.~\ref{fig:my_label6b}. Within the observable lattice spacing range, its envelope curve remains essentially distance-independent, showing no systematic decay toward zero beyond the NN or NNN range. Consistent with this, the corresponding magnetic response exhibits sharp rather than diffuse peaks in momentum space, corresponding to long-range order characteristics. This system realizes a long-range helical (spiral) magnetic state, although with weaker coherence than the N\'eel order, as reflected in the reduced real-space correlation amplitude and the broader momentum-space response.

\begin{figure}[H]
    \centering
    \includegraphics[width=0.5\textwidth]{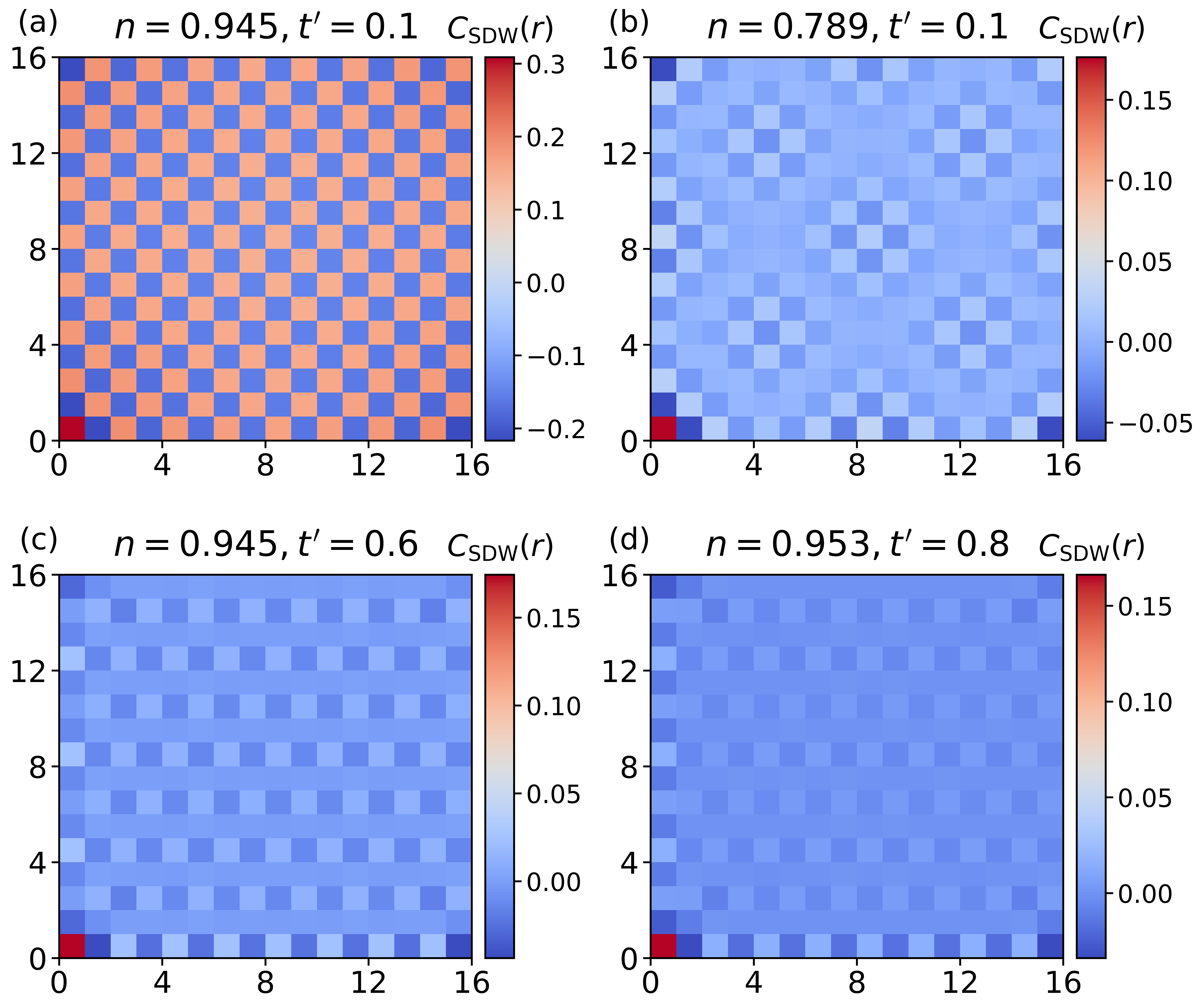}
    \caption{
The real space spin-spin correlation function $C_{\text{SDW}}(\mathbf{r})$ on a $16 \times 16$ lattice. (a) At $n=0.945$ and $t'=0.1$, the system exhibits a uniform checkerboard pattern, corresponding to a antiferromagnetic (Néel) order with ordering vector $\mathbf{Q}=(\pi, \pi)$. (b)-(d) With increased doping ($n=0.789$) or stronger frustration ($t'=0.6, 0.8$), the magnetic texture displays distinct spatial modulations, indicative of spiral magnetic order. Even for the doped case in (b), the correlation persist at long distances without decay, confirming the robustness of the long-range order.}
    \label{fig:my_label2}
\end{figure}

To further elucidate the nature of the magnetic order and verify its long-range coherence, we analyze the real space spin-spin correlation function. Figure~\ref{fig:my_label2} illustrates the spatial distribution of $C^z_{\text{SDW}}(\mathbf{r})$ on a $16 \times 16$ lattice for representative parameters. In the regime of weak frustration and near half-filling, as shown in Fig.~\ref{fig:my_label2}(a) ($n=0.945, t'=0.1$), the system exhibits a AFM order. The correlation function displays a uniform checkerboard pattern with alternating signs $(-1)^{r_x+r_y}$, corresponding to a Néel state with the ordering vector $\mathbf{Q}=(\pi, \pi)$. The amplitude of the correlation remains robust across the entire system size without observable decay, signifying a well-established long-range order. With increased doping or frustration, the magnetic texture evolves into spiral structures, as evidenced in Figs.~\ref{fig:my_label2}(b)-(d). The checkerboard pattern gives way to spatial modulations, indicating a shift of the ordering vector to positions $\mathbf{Q}=(\pi, q)$ or $(q, \pi)$. These patterns are signatures of spiral or stripe magnetic correlation driven by the competition between Fermi surface nesting and geometric frustration. Crucially, despite the spatial modulation, the envelope of the correlation function does not exhibit exponential decay at large distances ($|\mathbf{r}| \sim L/2$). The persistence of a sizable correlation amplitude at the largest separations confirms that these spiral states possess true long-range order rather than short-range liquid-like correlation. In reciprocal space, long-range magnetic order manifest as pronounced maxima in the spin structure factor at the corresponding ordering wave vectors, in contrast to the broad, diffuse features expected for short-range correlation. 
\subsection{C. Collinear stripe \((\pi, 0)\) and \((0, \pi)\) magnetic correlation}
\label{COLLINEAR MAGNETIC}
\begin{figure}[!htbp]
    \centering
    \includegraphics[width=0.5\textwidth]{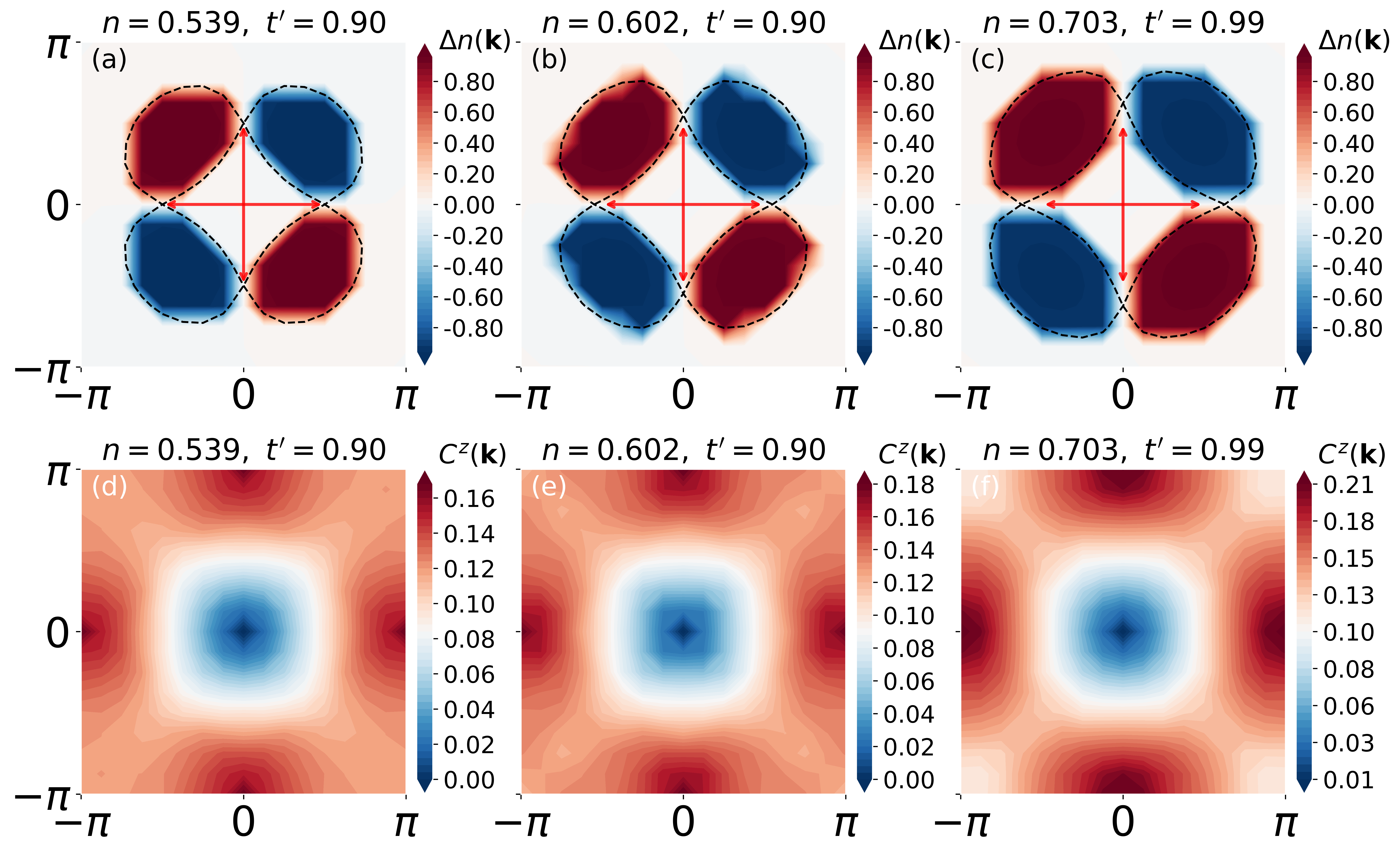}
    \caption{
The momentum-resolved spin imbalance and magnetic correlation response in the strong-$t'$ regime.
Top row: $\Delta n(\mathbf{k})$ overlaid with the corresponding noninteracting Fermi surfaces (dashed) and the dominant magnetic wave vector (red arrows) for
(a) $(n,t')=(0.539,0.90)$,
(b) $(n,t')=(0.602,0.90)$, and
(c) $(n,t')=(0.703,0.99)$.
Bottom row: corresponding $C^z(\mathbf k)$ for (d)–(f), showing that the dominant $C^z(\mathbf k)$ spectral weight is shifted toward axial ordering vector near $(\pi,0)$ and $(0,\pi)$, consistent with stripe magnetic correlation at large $t'$.}
    \label{fig:my_label5}
\end{figure}

For highly doped and large $t'$ regime, the ordering vector are no longer located at $ (\pi,\pi)$ but instead at the axial stripe ordering vector $\mathbf Q= (\pi,0)$ and $ (0,\pi)$. This change of magnetic correlation ordering vector can be understood already at the level of the noninteracting Fermi surface. The upper panels of Fig.~\ref{fig:my_label5} show the spin polarisation $\Delta n^\sigma (\mathbf k)$ for $t'\simeq 0.9$--0.99 and fillings $n=0.539,\,0.602,\,0.703$, together with the underlying Fermi surface (dashed lines). At these large $t'$, the Fermi surface is strongly distorted into four elongated pockets residing in each quadrant of the Brillouin zone. Importantly, the long sides of these pockets are oriented approximately parallel to the crystallographic axes, so that pairs of pockets on opposite sides of the Brillouin zone can be efficiently connected by purely axial ordering vector of the form $ (\pi,0)$ or $ (0,\pi)$, as indicated by the arrows in the top row. The shift by $\mathbf{Q}=(\pi,0)$ or $(0,\pi)$ yields only limited overlap between the relevant portions of the Fermi contour, indicating that the diagonal channel is only weakly nested.

From a weak-coupling viewpoint, the corresponding $C^z(\mathbf k)$ peak intensifies whenever a significant portion of the Fermi surface is nested by a given $\mathbf q$. For the large-$t'$ Fermi surfaces shown in Fig.~\ref{fig:my_label5}, the best nesting is achieved by the axial ordering vector $ (\pi,0)$ and $ (0,\pi)$ that connect the nearly parallel segments of opposite pockets. This is directly reflected in the lower panels of Fig.~\ref{fig:my_label5}, where the $C^z (\mathbf k)$ develops pronounced maxima along the Brillouin-zone edges at $\mathbf Q\approx (\pi,0)$ and $ (0,\pi)$ and shows no competing peak at $ (\pi,\pi)$. The evolution with filling in the three columns mainly changes the size and eccentricity of the pockets but does not alter the fact that the largest phase space for particle–hole scattering remains associated with axial, rather than diagonal, ordering vector, so that stripe magnetic correlation are stabilized over a broad density range.

This behavior can also be rationalized from a strong-coupling viewpoint, where the key physical ingredient is exchange frustration generated by the spin-dependent NNN hopping. In the limit $U\!\gg\! t,t'$, charge fluctuations are suppressed and the Hubbard model reduces, to leading order in $t/U$ and $t'/U$, to an effective $J_1$--$J_2$ Heisenberg model. NN hopping processes generate an antiferromagnetic superexchange $J_1\simeq 4t^2/U$, while NNN processes yield $J_2\simeq 4t'^2/U$. The competition between these two exchanges frustrates the N\'eel arrangement: $J_1$ favors antiparallel alignment between nearest neighbors and stabilizes $\mathbf{Q}=(\pi,\pi)$, whereas $J_2$ simultaneously favors antiparallel alignment across the diagonals and cannot be satisfied by the same N\'eel configuration on the square lattice. As $t'$ increases, $J_2/J_1$ grows rapidly and the frustration becomes sufficiently strong that the system minimizes its energy by reorganizing the spin pattern into a stripe state, in which one lattice direction is antiferromagnetic while the orthogonal direction becomes ferromagnetic \cite{chandra1990ising}. In the classical $J_1$--$J_2$ model this reorganization occurs once $J_2/J_1$ exceeds the well-known threshold $1/2$, beyond which the ground state switches from N\'eel order at $\mathbf{Q}=(\pi,\pi)$ to a collinear stripe manifold characterized by $\mathbf{Q}=(\pi,0)$ or $(0,\pi)$ \cite{jiang2012spin}. From this perspective, the dominant peaks in $C^z(\mathbf k)$ at large $t'$ can be understood as a shift of the ordering wave vector: increasing $t'$ strengthens competing diagonal exchange processes, and favors a reorganization of magnetic correlation into the collinear stripe patterns characterized by $\mathbf{Q}=(\pi,0)$ or $(0,\pi)$.

\begin{figure}[!htbp]
    \centering
    \includegraphics[width=0.5\textwidth]{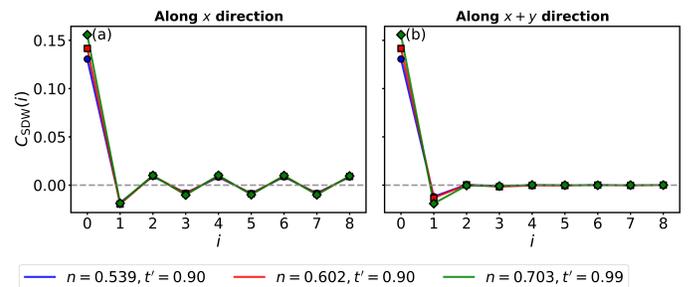}
    \caption{
The real space magnetic correlation function $C_{\mathrm{SDW}}(i)$ obtained from the inverse Fourier transform of $C^z(\mathbf k)$ for the same parameter sets as in Fig.~\ref{fig:my_label5}.
Panel (a) shows correlation along the $x$ direction, and panel (b) along the diagonal ($x{+}y$) direction.
In both panels, $C_{\mathrm{SDW}}(i)$ is dominated by an on-site component and rapidly decaying oscillations, indicating short-range magnetic correlation in the strong-$t'$ regime.}
    \label{fig:my_label5b}
\end{figure}

Similarly, we evaluate the real space magnetic correlation for a strongly spin splitting with $t'\sim 1$ over a range of electron fillings.  
As shown in Fig.~\ref{fig:my_label5b}, the resulting magnetic correlation patterns are qualitatively distinct from those at weak $t'$. For all fillings examined, the real space images display pronounced correlation only at the on-site and a few NNN separations, while the signal decays rapidly at larger distances. Despite the emergence of regular alternating patterns within the lattice, the values fluctuate minutely around zero with barely discernible correlation strength. This indicates that the N\'eel order is entirely disrupted, replaced by a short-range order corresponding to collinear stripe magnetism with finite correlation length. 

\subsection{D.  Coexistence regime of magnetic correlation}
\label{NON-COLLINEAR MAGNETIC}
\begin{figure}[H]
    \centering
    \includegraphics[width=0.5\textwidth]{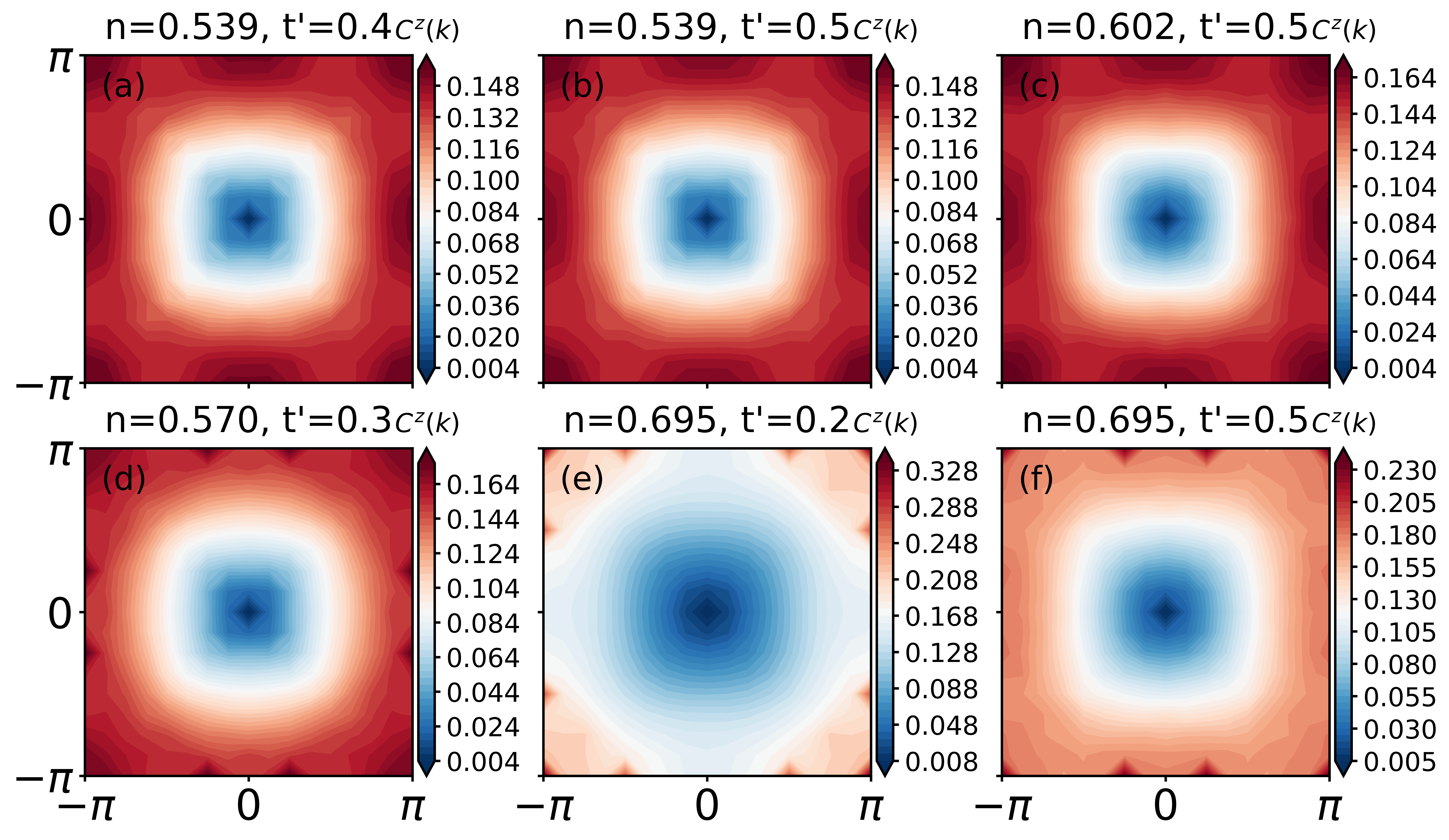}
    \caption{
Intensity distribution of the $C^z(\mathbf k)$ for different electron filling $n$ and next-nearest-neighbor transition strengths $t'$. (a)–(c) illustrate the competition and coexistence between collinear stripe correlation with ordering vector $\mathbf{Q}=(\pi, 0)/(0, \pi)$ and commutative N\'eel order with $\mathbf{Q}=(\pi, \pi)$. (d)–(f) depict the coexistence and competition mechanisms between the spiral magnetic correlation with the spiral correlation ordering vector $\mathbf{Q}=(\pi, q)/(q, \pi)$ and the N\'eel ordering vector with $\mathbf{Q}=(\pi, \pi)$. The color intensity in the figure represents the magnitude of the spectral weight.}
    \label{fig:my_label7}
\end{figure}

The momentum-resolved $C^z(\mathbf k)$ for various fillings $n$ and NNN hopping amplitudes $t'$, as shown in Fig.~\ref{fig:my_label7}. The evolution of the magnetic spectral weight reveals a complex phase diagram governed by the competition between N\'eel order ($\mathbf{Q}=(\pi, \pi)$), stripe correlation ($\mathbf{Q}=(\pi, 0)/(0, \pi)$), and spiral correlation ($\mathbf{Q}=(\pi, q)/(q, \pi)$).

The most pronounced competition occurs in the strong frustration regime, specifically at $(n, t')=(0.539, 0.4)$, $(n, t')=(0.539, 0.5)$ and $(0.602, 0.5)$ [Fig.~\ref{fig:my_label7}(a) - (b)]. Here, the system resides near a HO-VHS, where the Fermi surface undergoes a Lifshitz transition. This generates flattened bands near the antinodal points $(0, \pi)$ and $(\pi, 0)$, significantly enhancing the density of states and promoting the $\mathbf{Q}=(\pi, 0)/(0, \pi)$ ordering vector. However, the residual nesting of the curved Fermi surface segments continues to support the N\'eel ordering vector $\mathbf{Q}=(\pi, \pi)$. 
Consequently, the magnetic response does not lock into a single ordering vector. Instead, $C^z(\mathbf k)$ exhibits a broad, distribution of spectral weight covering the region between $(\pi, 0)$ and $(\pi, \pi)$. This signifies a robust coexistence and competition where the magnetic state fluctuates between diagonal N\'eel order and axial stripe correlation, driven by the degeneracy of the magnetic susceptibility at the ordering vector.

A complex coexistence regime emerges at intermediate frustration or higher doping levels, observed in $(n, t')=(0.570, 0.3)$, $(0.695, 0.2)$, and $(0.695, 0.5)$ [Fig.~\ref{fig:my_label7}(d)-(f)]. In these cases, the energy landscape is characterized by a subtle balance between the tendency to form spirals (driven by doping-induced nesting mismatches) and the corresponding N\'eel states persist under the influence of energy regulation.
The spectra display a composite structure: a distinct intensity maximum persists near the N\'eel ordering vector $\mathbf{Q}=(\pi, \pi)$, coexisting with broadened features or satellite peaks at the spiral correlation ordering vector $\mathbf{Q}=(\pi, q)/(q, \pi)$. Unlike Regime I, the component stripe correlation ordering vector $\mathbf{Q}=(\pi, 0)/(0, \pi)$ is subdominant or absent. This phenomenology suggests a state where the long-range order is frustrated, leading to a superposition of antiferromagnetic fluctuations and spiral modulations. The specific values of $t'$ and $n$ are used to adjust the relative weighting between these two components. Yet neither can completely suppress the other, resulting in a competitive mechanism.

To further assess whether the coexistence regime supports long-range order or remains short-ranged, we examine the momentum-space peak broadening of $C^z(\mathbf k)$. The width of the dominant peak provides a direct measure of magnetic coherence: a broadened maximum corresponds to diffuse spectral weight and is consistent with a finite dependent decay lengths, whereas a sharp and well-localized peak indicates coherent correlation extending across the system. For $n=0.695$ (both $t'=0.2$ and $0.5$), Fig.~\ref{fig:my_label7}(e,f) displays pronounced, sharply peaked features in $C^z(\mathbf k)$, signaling the development of long-range magnetic order. In contrast, for the other fillings investigated ($n=0.539, 0.570, 0.602$), panels (a--d) exhibit substantially broader peaks with appreciable momentum-space dispersion, indicating that the magnetic correlation remain short-ranged and do not develop global coherence. Consequently, outside the ordered phase at $n=0.695$, the system is dominated by short-range magnetic correlation, with quantum fluctuations preventing the symmetry breaking required for long-range order.

\section{\uppercase\expandafter{\romannumeral 4}. DISCUSSION AND CONCLUSION}

In this work, we have systematically investigated the magnetic properties of the two-dimensional Hubbard model, incorporating both NN and NNN hopping terms. Using CPQMC method, we researched and analyzed the system's magnetic order or correlation under varying doping concentrations \(n\) and NNN hopping \(t'\). Our results provide a comprehensive understanding of the evolution of magnetic order or correlation in this system, highlighting the interplay between  hopping parameters, magnetism distribution, and the resulting spin configurations.

By tuning the filling $n$ and the NNN hopping $t'$, the magnetic response evolves systematically, attributed to the modifications of the Fermi surface geometry and the  Fermi nesting. In the weak-$t'$ and low-doping regime ($n\simeq 1$), the $C^z(\mathbf k)$ develops a dominant and well-localized peak at $\mathbf{Q}=(\pi,\pi)$, indicating a strong propensity toward N\'eel antiferromagnetic order driven by near-perfect nesting. The peak exhibits low dispersion, and the real-space spin correlation order does not decay with distance, indicating long-range magnetic order.At near-half filling ($n\simeq 1$), increasing $t'$ progressively weakens the nesting channel and drives the $C^z(\mathbf k)$ response away from $\mathbf{Q}=(\pi,\pi)$ toward the ordering vector of the form $\mathbf{Q}=(\pi,q)$ or $(q,\pi)$. In this regime, the peak of $C^z(\mathbf k)$ remains relatively sharp over a finite range of $t'$, indicating that the correlation can retain long-range coherence.As $t'\to 1$ and $n\to 0.5$, the $C^z(\mathbf k)$ shifts toward the ordering vector of the form $\mathbf{Q}=(\pi,0)$ or $(0,\pi)$, consistent with stripe magnetic correlation. Concomitantly, the peak in the spin structure factor becomes markedly broadened, indicating that the magnetism is short-ranged with a finite correlation length. 

At intermediate fillings, the simultaneous degradation of the Fermi nesting and the redistribution of low-energy spectral weight further enhances the spiral response, with the $C^z(\mathbf k)$ centered at $\mathbf{Q}=(\pi,\pi)$ and $(\pi,q)$ or $(q,\pi)$. However, long-range coherence persists only within a narrow window of $t'$, and for $t'\gtrsim 0.5$ the peak becomes markedly broadened, signaling a crossover to short-range correlation with a finite correlation length. At lower fill factors ($n \sim 0.5$), the diagonal channel $(\pi,\pi)$ is strongly suppressed, while the axial scattering channel is relatively enhanced. As $t'$ increases, the ordering vector transitions from $\mathbf{Q}=(\pi,q)$ or $(q,\pi)$ to coexisting regime $\mathbf{Q}=(\pi,\pi)$ and $(\pi,0)$ or $(0,\pi)$, particularly near $t'\approx 0.5$. The proximity to the Van Hove singularity expands the available phase space for axial particle-hole scattering, thereby promoting stripe correlation characterized by $\mathbf{Q}=(\pi,0)$ or $(0,\pi)$. The corresponding peak in $C^z(\mathbf k)$ is relatively diffuse, indicating short-range spin correlation within this parameter region. 
These observations imply that the intermediate coexistence region is not a static mixture of distinct phases, but rather a competition-dominated regime arising from frustrated and nesting-controlled ordering channels. As a result, the system does not lock into a unique ordering vector: at lower $n$ it exhibits pronounced stripe--N\'eel competition, whereas at intermediate fillings it is governed by spiral-N\'eel competition.

In conclusion, our research demonstrates the intricate relationship between electron interactions, hopping terms, and the resulting magnetic phases in the two-dimensional Hubbard model. The transition from collinear N\'eel magnetic order to non-collinear magnetic correlation, and eventually to spiral configurations, is driven by the momentum-dependent effects of the NNN hopping \(t'\), highlighting the importance of NNN interactions in stabilizing complex magnetic correlation. Our results contribute to a deeper understanding of the magnetic phase diagram of correlated electron systems and open avenues for further investigation into the role of NNN interactions in quantum magnetic materials.

These results may also help motivate materials-oriented studies of systems proposed to exhibit a d-wave--like altermagnetic spin splitting. In particular, materials such as RuO$_2$~\cite{ruo2_2025prb}, KV$_2$Se$_2$O~\cite{jiang2025kv2se2o}, and La$_2$O$_3$Mn$_2$Se$_2$
~\cite{wei2025la2o3mn2se2} are often discussed as candidates in this context. From a device viewpoint, such tunability of the magnetic correlation provides a natural handle for spintronics in compensated settings. In particular, the evolving magnetic texture can affect the response of anisotropic magnetoresistance and Hall effect demonstrated in altermagnetic MnTe~\cite{betancourt2024amr_mnte}.

\appendix
\section*{Appendix A: Finite-Size Dependence of spin structure factor Peak Locations}

We compare the dominant $C^z(\mathbf k)$ peak locations across the $(n,t')$ plane for $L=18,14,12,$ and $8$ to assess finite-size effects. The overall distribution of the peak locations is robust: the same sequence of dominant ordering tendencies upon varying $n$ and $t'$ persists as $L$ is changed. The most visible size dependence occurs in the spiral sector, where the allowed momenta are discretized on an $L\times L$ grid. As the lattice size increases, the distribution of magnetic ordering tendencies becomes progressively smoother and more continuous in parameter space, while the overall structure of the peak pattern remains essentially unchanged.
Regions where two nearby maxima have comparable intensity are more sensitive to $L$, since a modest change in momentum resolution can shift which peak is dominant. For larger lattices ($L=12$--$18$), this sensitivity is reduced and the boundaries become more stable, supporting the robustness of our conclusions against finite-size effects.

\begin{figure}[!htbp]
    \centering
    \includegraphics[width=0.5\textwidth]{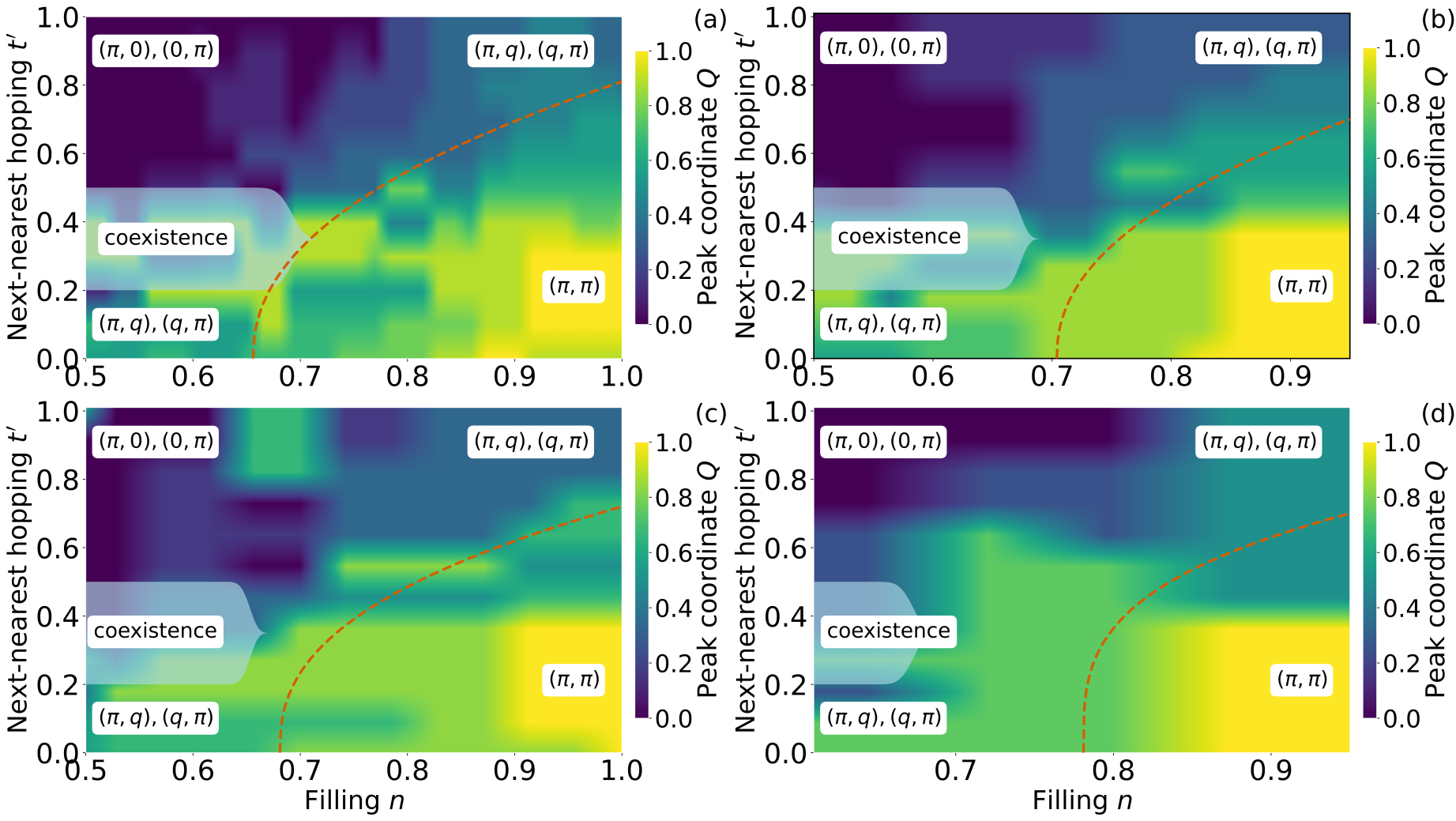}
    \caption{
The $C^z(\mathbf k)$ peak phase diagrams in the $(n,t')$ plane for (a) $L=18$, (b) $L=14$, (c) $L=12$, and (d) $L=8$.
Each panel shows the $C^z(\mathbf k)$ peak coordinate $Q$ at which $|C^z(\mathbf k)|$ is maximal:
$Q=1$ corresponds to N\'eel order at $(\pi,\pi)$, $Q=0$ to collinear stripe correlation at
$(\pi,0)$ or $(0,\pi)$, and $0<Q<1$ indicates spiral $C^z(\mathbf k)$ peaks
[e.g., $(\pi,q)$ or $(q,\pi)$]. Long and short order for magnetic correlation are distinguished by red lines. Regions to the right of the red line represent long-range order, while those to the left indicate short-range order.
}
    \label{fig:appendix_1}
\end{figure}

\section*{Appendix B: Distribution of VHS across different $t'$}
\begin{figure}[!htbp]
    \centering
    \includegraphics[width=0.5\textwidth]{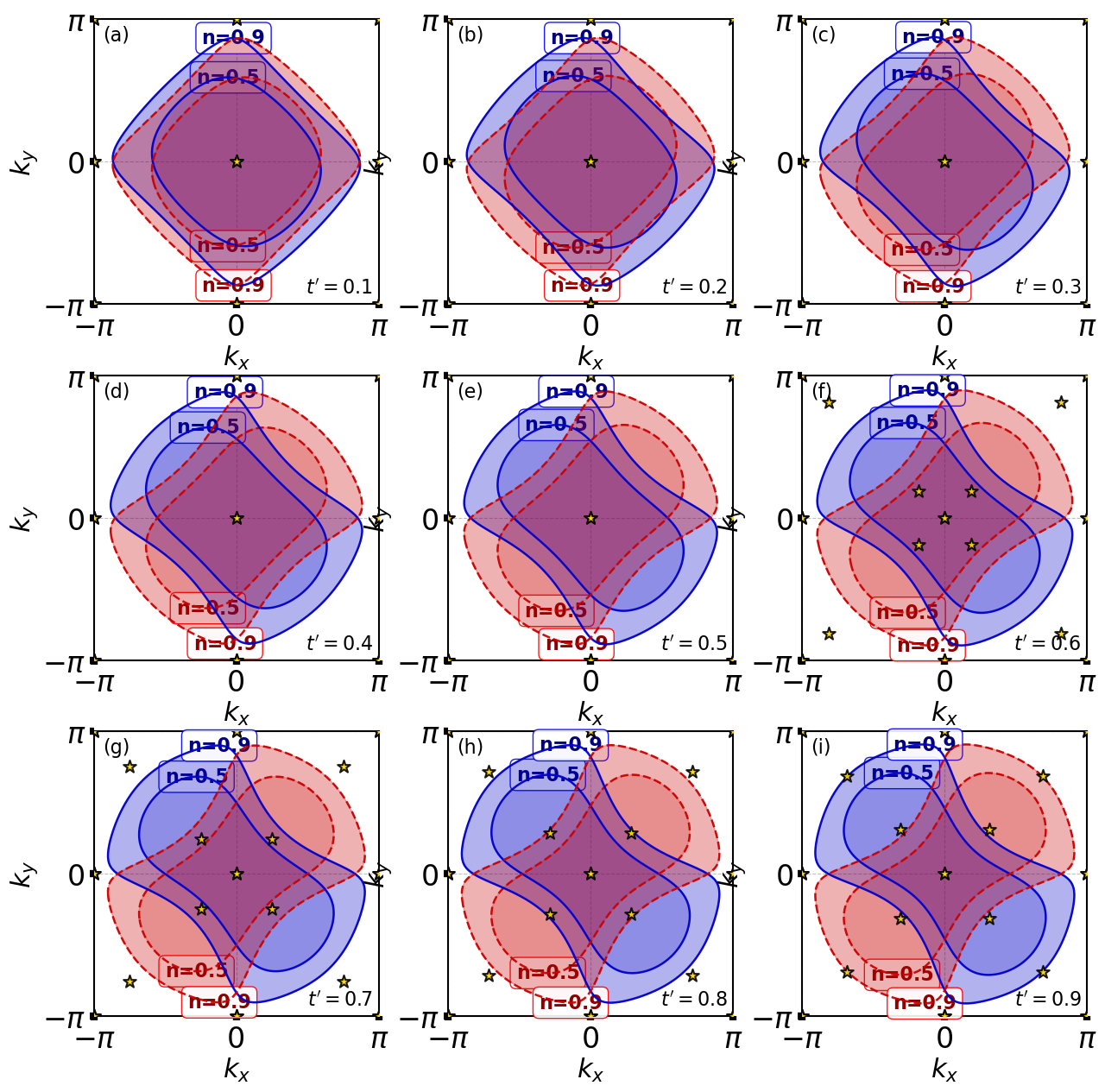}
    \caption{The evolution of VHS and Fermi surfaces with varying NNN hopping $t'$. 
Panels (a)–(i) correspond to $t' = 0.1, 0.2, \ldots, 0.9$. For each $t'$, solid (dashed) contours denote the Fermi surfaces of the upper (lower) band at fillings $n = 0.5$ and $n = 0.9$, while the shaded regions indicate the corresponding filled Fermi surface.
Stars mark momenta where $\nabla_{\mathbf{k}}\varepsilon_{\mathbf{k}}=0$, i.e., VHS saddle points. As $t'$ increases, the VHS move away from the high-symmetry points and the Fermi surfaces become strongly distorted, degrading $ (\pi,\pi)$ nesting and thereby favoring $ ( \pi, q)$ or $ ( q, \pi)$ spiral tendencies at larger $t'$.}
    \label{fig:appendix_2}
\end{figure}

We present the distribution of VHS in momentum space for different values of the spin–dependent hopping amplitude $t'$.  For each $t'$, we determine the saddle points of the band dispersion and track their locations in the Brillouin zone. The resulting trajectories are shown in Fig.~\ref{fig:appendix_2}, where the yellow (green) curves denote the VHS associated with the lower (upper) band.

A clear qualitative change occurs around a characteristic value $t' \simeq 0.5$.  
For $t' < 0.5$, the VHS remain pinned to the high–symmetry points $Y (0,\pi)$, $X (\pi,0)$, $\Gamma (0,0)$, and $M (\pi,\pi)$, indicating that the dominant saddle points are commensurate with the underlying lattice.  In contrast, for $t' >0.5$ the VHS detach from these high–symmetry points and migrate onto the diagonals $k_x = \pm k_y$, giving rise to spiral saddle points.  As $t'$ is further increased, the VHS trajectories spread out along these diagonals and the corresponding saddle energies become more strongly dispersed, reflecting the enhanced anisotropy of the band structure induced by $t'$.

This systematic evolution of the VHS distribution encapsulates the way in which the NNN hopping $t'$ reshapes the Fermi–surface and modifies the nesting conditions. In particular, the relocation of the VHS from high–symmetry points to spiral momenta directly impacts the dominant magnetic instabilities.  Combined with the main–text analysis of the $C^z(\mathbf k)$ response, these results demonstrate that tuning $t'$ provides an efficient handle to control both the position and the spectral weight of the VHS, thereby governing the evolution of the system's magnetic correlation tendencies.

\textcolor{black}{\section*{Appendix C: magnetic correlation in a non-interacting system}}

\begin{figure}[!htbp]
    \centering
    \includegraphics[width=0.5\textwidth]{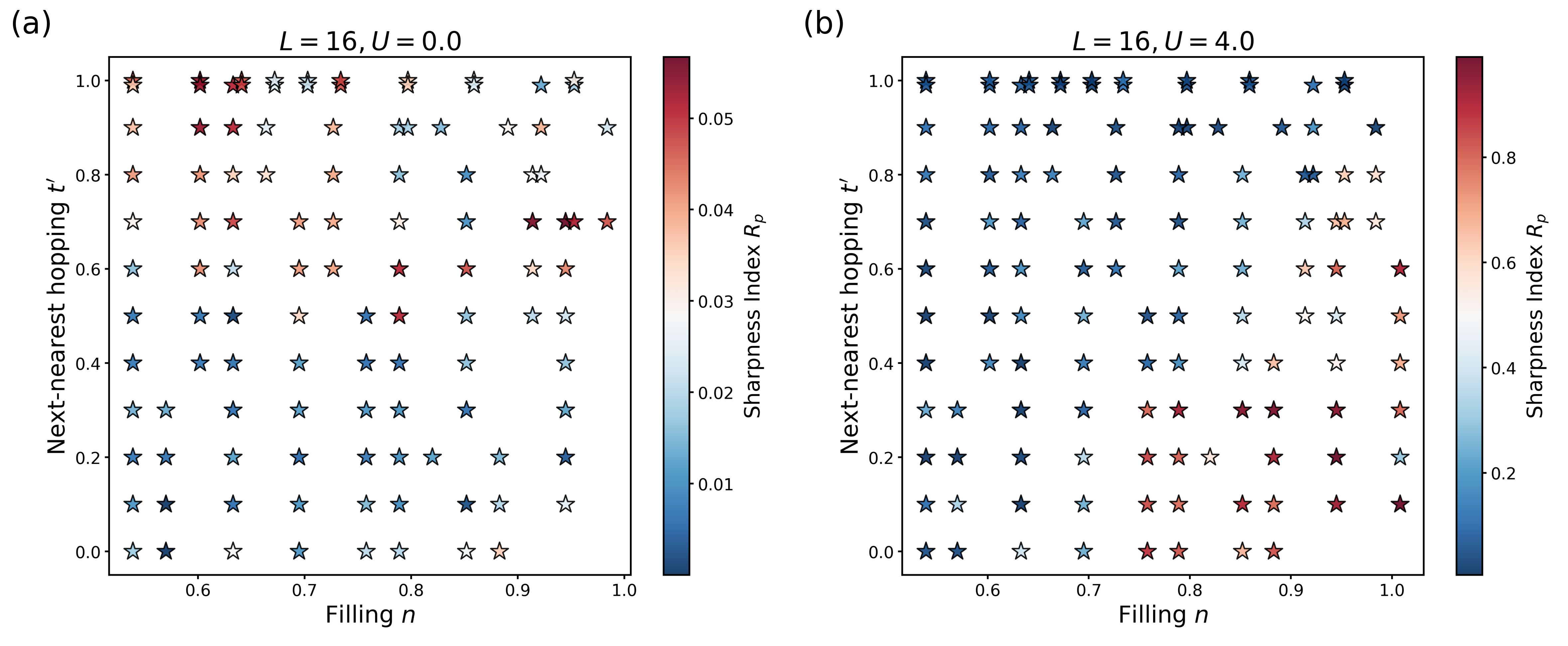}
    \caption{\textcolor{black}{The dispersion distributions of the $C^z(\mathbf k)$ peak correspond to: (a) $U=0$, (b) $U=4$. In the subgraph, $R_p$ denotes the sharpness of the $C^z(\mathbf k)$ peak, where a larger $R_p$ value indicates more stable magnetic ordering in $C^z(\mathbf k)$:
$R_p$ is generally low at $U=0$, while $R_p$ significantly increases at $U=4$ for larger $n$ and smaller $t'$. In this region, magnetic correlations exhibit long-range order. For $n \geq 0.539$ and $t' \geq 0.5$ where both parameters increase, the relative change in $R_p$ between $U=0$ and $U=4$ is small.}}
    \label{fig:appendix_3}
\end{figure}

\textcolor{black}{In the noninteracting limit ($U=0$), the model is a quadratic free-fermion problem with a spin-dependent dispersion. In this case, the dominant factors governing magnetic correlation are primarily determined by the geometry of the Fermi surface and the particle-hole phase space regulated by VHS, and there is no interaction-driven mechanism to generate collective spin order or spontaneous symmetry breaking. Accordingly, the momentum-space spin response is broad and diffuse, and the corresponding $C^z(\mathbf k)$ pattern should be interpreted as reflecting the bare polarization function $\chi_{0}(\mathbf{q})$, rather than a true ordered state with a long correlation length. Even when the Fermi surface exhibits favorable nesting at particular wave vectors, the absence of an interaction scale prevents these fluctuations from being locked in and amplified into stable long-range correlations, as shown in Fig.~\ref{fig:appendix_3}.} \textcolor{black}{To quantify the parameter dependence of spatial coherence, we introduce a single scalar peak-sharpness metric $R_p(n,t')=(C_{\max}-\langle C_{\mathrm{near}}\rangle)/C_{\max}$, which measures how sharply $C^z(\mathbf{k})$ is peaked. The resulting $R_p$ map over the $(n,t')$ plane is shown in Fig.~\ref{fig:appendix_3}.}

\textcolor{black}{Upon increasing $t'$ into the regime $t'\ge 0.5$ and  $n \ge 0.539$, the single-particle band geometry becomes the primary control knob for the dominant scattering channels, as shown in Fig.~\ref{fig:appendix_3_1}. In this itinerant regime, the magnetic wave vector selection in $C^z(\mathbf k)$ is largely set by band-structure kinematics; turning on interactions mainly enhances the leading channel without substantially shifting its location. This explains why the overall momentum-space patterns for $U=0$ and $U=4$ become increasingly similar as $t'$ and $n$ increase.
\begin{figure}[!htbp]
    \centering
    \includegraphics[width=0.5\textwidth]{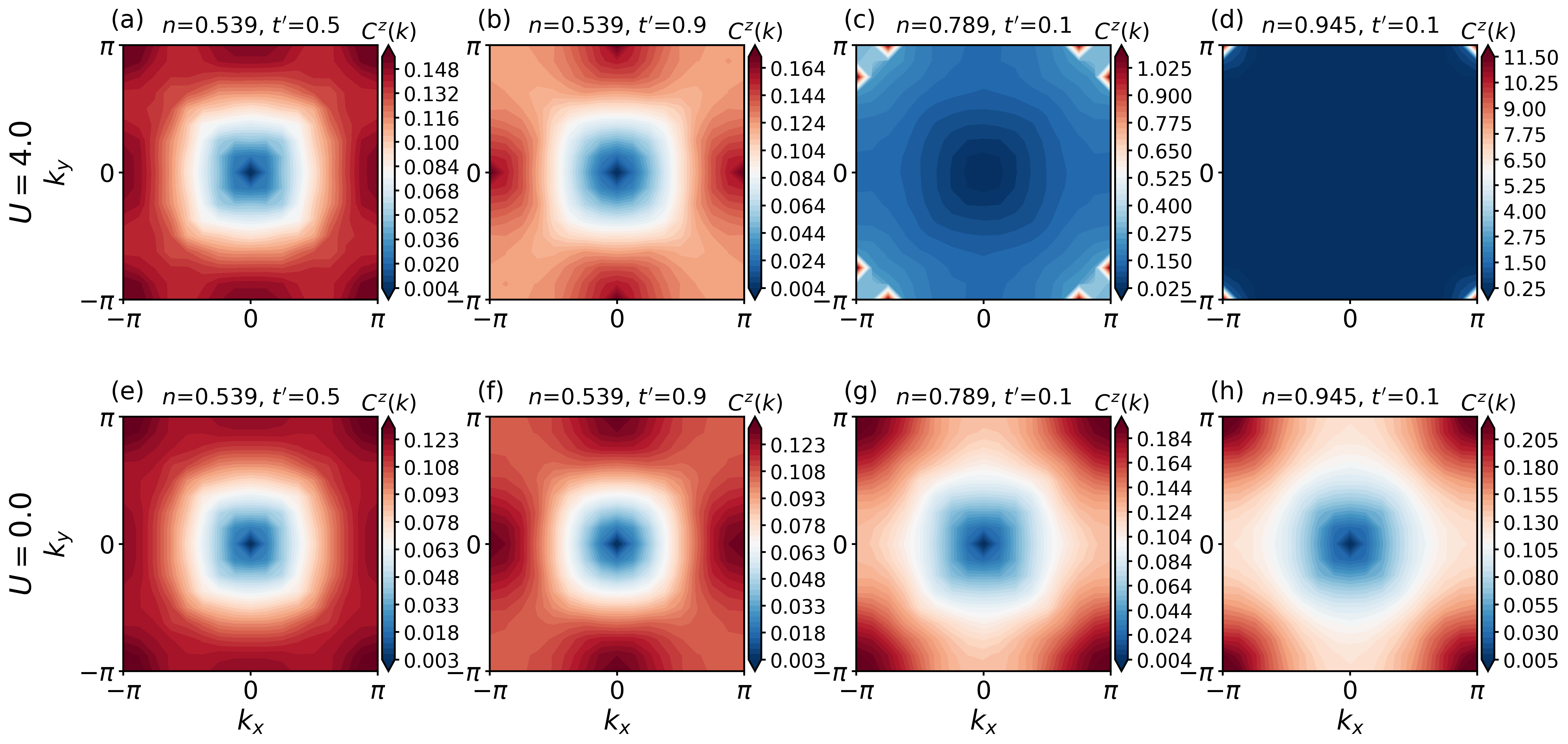}
    \caption{\textcolor{black}{Intensity distributions of $C^z(\mathbf k)$ for different electron filling $n$ and next-nearest-neighbor coupling strength $t'$ at $U=0$ and $U=4$. (a)–(d) illustrate distinct magnetic correlation features at $U=4$, including coexistence, spiral, and Neel phases. (d)–(f) depict the magnetic correlation distributions for $U=0$. It is observed that at $n=0.539$, the $C^z(\mathbf k)$ images for $U=0$ and $U=4$ are similar. When $n=0.789$ and $n=0.945$, and $t'=0.1$, $U=4$ exhibits spiral magnetic order and Neel antiferromagnetic order, respectively, showing significant differences from $U=0$. The color intensity in the figure represents the magnitude of spectral weight.}}
    \label{fig:appendix_3_1}
\end{figure}
In contrast, at finite interaction strength ($U=4$) and especially for smaller $t'$ as illustrated Fig.~\ref{fig:appendix_3_1}, the system enters a correlation-dominated regime where local-moment formation and superexchange-driven competition (including frustration effects akin to $J_{1}$--$J_{2}$ model) qualitatively reshape the spin response. The resulting $C^z(\mathbf k)$ peak becomes much sharper than at $U=0$, signaling a substantially increased magnetic correlation length and a greater propensity toward ordered or quasi-ordered magnetism. Within this framework, the absence of a pronounced spiral texture at $U=0$ is natural: spiral correlations require interaction-driven enhancement of an incommensurate channel (through vertex corrections and many-body feedback), which is not present in the noninteracting response. At $U=4$, such many-body effects can selectively amplify an incommensurate wave vector while suppressing competing channels, thereby producing the well-defined spiral peak observed in $C^z(\mathbf k)$.}

\FloatBarrier

\bibliography{references}
\end{document}